# Spectrum of Low-Lying Excitations in a Supersymmetric Extended Hubbard Model [1]

Fabian H.L. Eßler [b]

and

Vladimir E. Korepin [♯]

*Institute for Theoretical Physics*
*State University of New York at Stony Brook*
*Stony Brook, NY 11794-3840*

## ABSTRACT

We continue the study of the $u(2|2)$-supersymmetric extension of the Hubbard model in one dimension. We determine the excitation spectrum at zero temperature even in the sectors where the ground states are $u(2|2)$-descendants of Bethe states. The excitations include spinons, holons, electrons, localons (local electrons pairs, moving coherently through the lattice) and their bound states. The spectra are found to be very different for repulsive and attractive on-site interaction. We also study the thermodynamics of the model.

---

[1] This work was supported in part by NATO, S.P. Chaos, Order and Pattern (CRG 901098).
[b] E-MAIL: FABMAN@MAX.PHYSICS.SUNYSB.EDU
[♯] E-MAIL: KOREPIN@MAX.PHYSICS.SUNYSB.EDU

## 1. Introduction

This is the fourth part of a series of papers dealing with the $u(2|2)$-symmetric model of superconductivity. In [1] a new model of strongly correlated electrons on a general $d$-dimensional lattice was proposed, which differs from the Hubbard model by moderate interactions of nearest neighbours. The additional interactions fall into two categories : the so-called "bond-charge repulsion" terms and nearest neighbour spin-spin and charge interactions are similar to the ones of the model of hole superconductivity proposed in [2], and the $t$-$J$ model respectively. The pair-hopping term (which allows local electron pairs to move coherently) has been studied extensively in the Penson-Kolb-Hubbard models[3-5]. The model is of the "permutation type" first studied by Sutherland[6]. The model has the interesting property that the ground states for attractive and moderately repulsive on-site interaction exhibit off-diagonal long-range order (ODLRO) and are thus superconducting in one, two and three dimensions[7-9]. Furthermore it is possible to derive the exact expression for the unique ground state in the attractive regime in any number of dimensions. In one dimension the model is integrable and in [10] the nested Algebraic Bethe Ansatz solutions were constructed in detail. That paper also contains expressions for higher conservation laws and the proof of a lowest weight theorem of the Bethe Ansatz states with respect to the $u(2|2)$ symmetry algebra of the model.

This paper is a direct continuation of [10]. Starting from the Bethe Ansatz equations in the BBFF representation[2] derived in [10] (see also [11]) we first derive the infinite set of coupled integral equations that drive the thermodynamics of the model (TBA equations). By means of taking the zero temperature limit of these equations, we determine the ground state of the model in one dimension. The result agrees perfectly with the result obtained in [10] by completely different methods. The zero temperature limit of the TBA equations also provides expressions for the dressed energies of the elementary excitations that are given by the Bethe Ansatz. We also study the limit of infinite temperature of the TBA equations. In section 4 we determine the spectrum of low lying excitations in the $T = 0$ formalism. We carefully take into account that the Bethe Ansatz only provides lowest weight states of the $u(2|2)$ symmetry algebra[10]. This makes it necessary to analyze not only the Bethe equations in the thermodynamic limit, but also to construct the complete $u(2|2)$ multiplet from the lowest weight state obtained by means of the Bethe Ansatz. In order to enumerate all elementary excitations we use *both* Bethe Ansatz and global symmetry (see also [12-14]). A complete set of elementary excitations is of course necessary for determining questions like charge and spin separation, or deriving effective field theories for the model. The results for *repulsive* on-site interaction (and filling greater than zero) fare as follows: Apart from spinons and holons, which are already present in the supersymmetric $t$-$J$ model (which is a submodel of the $u(2|2)$-model) with linear dispersion relations around their repective Fermi surfaces, there exist electronic excitations with quadratic dispersions and a gap, and excitations involving local electron pairs. These also have a quadratic dispersion and a gap for $\mu < 0$ ($\mu$ is the chemical potential), but become gapless for $\mu = 0$. For *attractive* on-site interaction we find an infinite number of excitations with cosine-like dispersions. Amongst these are electrons and local pairs and their bound states.

---

[2] The reason for choosing the BBFF representation is twofold : First it reduces to the Sutherland representation of the supersymmetric $t$-$J$ model in the sector without localons. Second the ground state derived below is of a very simple form - there are no complex roots of the Bethe equations present in the ground state.

The hamiltonian on a lattice of $L$ sites in the grand canonical ensemble is[1]

$$H(\mu, U, h) = H^0 + U \sum_{j=1}^{L} (n_{j,1} - \frac{1}{2})(n_{j,-1} - \frac{1}{2}) - \mu \sum_{j=1}^{L} n_j + h \sum_{j=1}^{L} (n_{j,1} - n_{j,-1}) \quad (1.1)$$

where $H^0 = -\sum_{j=1}^{L} H^0_{j,j+1}$ is given by

$$\begin{aligned}
H^0_{j,j+1} &= c^\dagger_{j+1,1} c_{j,1} (1 - n_{j,-1} - n_{j+1,-1}) + c^\dagger_{j,1} c_{j+1,1} (1 - n_{j,-1} - n_{j+1,-1}) \\
&+ c^\dagger_{j+1,-1} c_{j,-1} (1 - n_{j,1} - n_{j+1,1}) + c^\dagger_{j,-1} c_{j+1,-1} (1 - n_{j,1} - n_{j+1,1}) \\
&+ \frac{1}{2}(n_j - 1)(n_{j+1} - 1) + c^\dagger_{j,1} c^\dagger_{j,-1} c_{j+1,-1} c_{j+1,1} + c_{j,-1} c_{j,1} c^\dagger_{j+1,1} c^\dagger_{j+1,-1} \\
&- \frac{1}{2}(n_{j,1} - n_{j,-1})(n_{j+1,1} - n_{j+1,-1}) - c^\dagger_{j,-1} c_{j,1} c^\dagger_{j+1,1} c_{j+1,-1} - c^\dagger_{j,1} c_{j,-1} c^\dagger_{j+1,-1} c_{j+1,1} \\
&+ (n_{j,1} - \frac{1}{2})(n_{j,-1} - \frac{1}{2}) + (n_{j+1,1} - \frac{1}{2})(n_{j+1,-1} - \frac{1}{2}) \, .
\end{aligned}$$
(1.2)

Here $U$ is the Hubbard model coupling constant, $h$ is an external magnetic field, and $\mu$ is the chemical potential. By $n_{i,\sigma} = c^\dagger_{i,\sigma} c_{i,\sigma}$ we denote the number operator for electrons with spin $\sigma$ on site $i$ and we write $n_i = n_{i,1} + n_{i,-1}$. Under the particle-hole transformation $c^\dagger_{j,\sigma} \leftrightarrow c_{j,\sigma}$ the hamiltonian $H(\mu, U, h)$ transforms (up to a constant) into $H(-\mu, U, h)$ so that it is sufficient to consider only the region $\mu \leq 0$ of the chemical potential. It was shown in [1] that $H^0$ is invariant under a $u(2|2)$ symmetry algebra. The three other terms in $H$ are elements of the Cartan subalgebra of $u(2|2)$ (and commute with $H^0$ and each other), and thus $H^0$ and $H$ have a complete set of simultaneous eigenstates. The dynamics of the model are such that the individual numbers $N_\uparrow$ and $N_\downarrow$ of electrons with spin up and spin down, and the numbers $N_l$ and $N_h$ of doubly occupied ("local electron pairs") and empty sites ("holes") are conserved quantities. We will choose the following conventions throughout this paper

$N_\uparrow$ = number of single electrons with spin up
$N_\downarrow$ = number of single electrons with spin down
$N_e = N_\uparrow + N_\downarrow$ = number of single electrons
$N_l$ = number of local electron pairs

$N_h$ = number of holes
$N_b = N_h + N_l$ = number of "bosons".

Let us now review a few facts about the $u(2|2)$ symmetry of the model[3], which we will need in section 4 below. There are a total of 16 generators, 8 of which are fermionic and bosonic respectively. The eight bosonic operators fall into two $su(2)$ and two $u(1)$ subalgebras. The two $su(2)$ algebras are generated by the spin-operators

$$S = \sum_{j=1}^{L} c^\dagger_{j,1} c_{j,-1} , \quad S^\dagger = \sum_{j=1}^{L} c^\dagger_{j,-1} c_{j,1} , \quad S^z = \sum_{j=1}^{L} \frac{1}{2}(n_{j,1} - n_{j,-1}) , \qquad (1.3)$$

and by the $\eta$-pairing like operators[15]

$$\eta = \sum_{j=1}^{L} c_{j,1} c_{j,-1} , \quad \eta^\dagger = \sum_{j=1}^{L} c^\dagger_{j,-1} c^\dagger_{j,1} , \quad \eta^z = \sum_{j=1}^{L} -\frac{1}{2} n_j + \frac{1}{2} . \qquad (1.4)$$

The two $u(1)$ charges are given by the identity operator and the Hubbard interaction

$$X = \sum_{j=1}^{L} (n_{j,1} - \frac{1}{2})(n_{j,-1} - \frac{1}{2}) . \qquad (1.5)$$

The eight fermionic generators are given by

$$Q_\sigma = \sum_{j=1}^{L} (1 - n_{j,-\sigma}) c_{j,\sigma} , \quad \tilde{Q}_\sigma = \sum_{j=1}^{L} n_{j,-\sigma} c_{j,\sigma} , \quad \sigma = \pm 1 \qquad (1.6)$$

and their hermitean conjugates. The lowest weight theorem states that all states constructed by Bethe Ansatz are lowest weight states of the $u(2|2)$ symmetry algebra[10]. In the BBFF representation of the Bethe Ansatz this implies that ($|BA\rangle$ denotes a Bethe state)

$$\eta |BA\rangle = 0 = S|BA\rangle = Q^\dagger_\sigma |BA\rangle = \tilde{Q}_\sigma |BA\rangle .$$

A complete set of states can be obtained by acting with the $u(2|2)$ raising operators on the Bethe states. Every Bethe state gives rise to one $u(2|2)$ multiplet. These lowest weight representations fall into two categories: so-called typical and atypical ones[16,17]. The dimensions of the typical representations are given by

$$\dim(N_\uparrow, N_\downarrow, N_l)\Big|_{typ} = 16(N_\uparrow - N_\downarrow + 1)(L - N_\uparrow - N_\downarrow - 2N_l + 1) , \qquad (1.7)$$

where $N_\uparrow - N_\downarrow + 1$ and $L - N_\uparrow - N_\downarrow - 2N_l + 1$ are the dimensions of the representaions of the spin and $\eta$-pairing $su(2)$ algebras respectively, and the factor of 16 is due to all possible actions of the fermionic generators. For the atypical representations some (nontrivial) actions of the fermionic generators give zero, and the multiplets are smaller. Two such cases appear for the model at hand

$$\dim(L, 0, 0) = 4L,$$
$$\dim(N_\uparrow, 1, 0) = \dim(N_\uparrow - 1, 0, 1) = 8(N_\uparrow + N_l - 1)(L - N_\uparrow - N_l) + 4L .$$
(1.8)

---
[3]More precisely of $H^0$, as only $H^0$ commutes with all generators of the $u(2|2)$ algebra.

Whereas (1.2) commutes with all 16 generators of $u(2|2)$, the hamiltonian (1.1) does not as the chemical potential, magnetic field, and the Hubbard interaction break the symmetry in a trivial way. The energies of the $u(2|2)$ descendants of the Bethe states $|BA\rangle$ are shifted as follows (for zero magnetic field)

$$H(\mu, U, 0) \ |BA\rangle = E \ |BA\rangle$$
$$H(\mu, U, 0) \ S^\dagger \ |BA\rangle = E \ S^\dagger \ |BA\rangle$$
$$H(\mu, U, 0) \ \eta^\dagger \ |BA\rangle = (E - 2\mu) \ \eta^\dagger \ |BA\rangle$$
$$H(\mu, U, 0) \ \tilde{Q}^\dagger_\sigma \ |BA\rangle = (E + \frac{U}{2} - \mu) \ \tilde{Q}^\dagger_\sigma \ |BA\rangle$$
$$H(\mu, U, 0) \ Q_\sigma \ |BA\rangle = (E + \frac{U}{2} + \mu) \ Q_\sigma \ |BA\rangle .$$
(1.9)

Equations (1.9) will be important for the classification of the excitation spectrum in section 4.

## 2. String Hypothesis and Logarithmic Form of the Bethe Equations

In this section we classify all roots of the Bethe equations in the BBFF representation[10,11] by means of a usual "string picture". After inserting this classification into the Bethe equations and taking the logarithm, solutions of the Bethe equations are unambiguously parametrized by a set of integer of half-odd integer numbers. This is the starting point for both the Thermodynamic Bethe Ansatz constructed in section 3, and the determination of the excitation spectrum at zero temperature (section 4).

The Bethe Ansatz equations for the BBFF grading are of the form

$$\left(\frac{\lambda_k - i}{\lambda_k + i}\right)^L = \prod_{j=1}^{N_b} \frac{\lambda_k - \lambda_j^{(1)} + i}{\lambda_k - \lambda_j^{(1)} - i} \prod_{\substack{l=1 \\ l \neq k}}^{N_b + N_\downarrow} \frac{\lambda_l - \lambda_k + 2i}{\lambda_l - \lambda_k - 2i} \quad k = 1, \ldots, N_b + N_\downarrow$$

$$\prod_{k=1}^{N_\downarrow + N_b} \frac{\lambda_k - \lambda_j^{(1)} + i}{\lambda_k - \lambda_j^{(1)} - i} = \prod_{m=1}^{N_l} \frac{\lambda_m^{(2)} - \lambda_j^{(1)} + i}{\lambda_m^{(2)} - \lambda_j^{(1)} - i} \qquad j = 1, \ldots, N_b ,$$
(2.1)

$$\prod_{\substack{l=1 \\ l \neq m}}^{N_l} \frac{\lambda_l^{(2)} - \lambda_m^{(2)} + 2i}{\lambda_l^{(2)} - \lambda_m^{(2)} - 2i} = \prod_{j=1}^{N_b} \frac{\lambda_m^{(2)} - \lambda_j^{(1)} - i}{\lambda_m^{(2)} - \lambda_j^{(1)} + i} \qquad , \qquad m = 1, \ldots, N_l .$$

Here we have rescaled the spectral parameters by a factor of 2 as compared to formula (2.78) in [10]. Energy and momentum of a state parametrized by a solution of (2.1) are given by (in the grand canonical ensemble)

$$E(\lambda_1,\ldots,\lambda_n) = L - \sum_{j=1}^{N_b+N_\downarrow} \frac{4}{\lambda_j^2+1} - \mu\left(N_\downarrow+N_\uparrow+2N_l\right) + h\left(N_\uparrow-N_\downarrow\right) + U\left(N_l - \frac{N_e}{2} + \frac{L}{4}\right) \ ,$$

$$p(\lambda_1,\ldots,\lambda_n) = (L+1)\pi - i\sum_{j=1}^{N_b+N_\downarrow} \ln\left(\frac{\lambda_j-i}{\lambda_j+i}\right) \ .$$
(2.2)

In the limit of large lattice lenghts $L$ the imaginary parts of solutions of (2.1) are distributed according to the following idealized presciption (string hypothesis)

$$\lambda\text{-strings} : \quad \lambda_\alpha^{nj} = \lambda_\alpha^n + i(n+1-2j) \ , \lambda_\alpha^n \in \mathbb{R} \ , \quad j = 1\ldots n \ , \tag{2.3}$$

$$\text{real } \lambda^{(1)} : \quad \lambda_\gamma^{(1)} \in \mathbb{R} \ , \tag{2.4}$$

$$\lambda^{(1)}-\lambda^{(2)}\text{-strings} : \begin{cases} \lambda_\gamma^{(1)\,2j} = \lambda_\gamma^{(2)} + i(3-2j) \ , & j=1,2 \\ \lambda_\gamma^{(2)} \in \mathbb{R} \end{cases} , \tag{2.5}$$

$$\lambda^{(2)}\text{-strings} : \quad \lambda_\nu^{(2)\,pk} = \lambda_\nu^{(2)\,P} + i(p+1-2k) \ , \lambda_\nu^{(2)\,P} \in \mathbb{R}, \quad k=1\ldots p \ . \tag{2.6}$$

Inserting (2.3)-(2.6) into (2.1) and "multiplying out the strings" we arrive at the following equations involving only the real centers of the strings

$$e^L\left(\frac{\lambda_\alpha^n}{n}\right) = \left(\prod_{(\beta,m)\neq(\alpha,n)} E_{mn}\left(\lambda_\alpha^n - \lambda_\beta^m\right)\right) \prod_{\gamma=1}^{M_1^{(1)}} e\left(\frac{\lambda_\gamma^{(1)\,1} - \lambda_\alpha^n}{n}\right) \prod_{\delta=1}^{M_2^{(1)}} E_{n1}\left(\lambda_\delta^{(1)\,2} - \lambda_\alpha^n\right)$$

$$\prod_{n=1}^{\infty}\prod_{\alpha=1}^{M_n} e\left(\frac{\lambda_\gamma^{(1)\,1} - \lambda_\alpha^n}{n}\right) = \prod_{p=1}^{\infty}\prod_{\nu=1}^{M_p^{(2)}} e\left(\frac{\lambda_\gamma^{(1)\,1} - \lambda_\nu^{(2)\,P}}{p}\right) \prod_{\delta=1}^{M_2^{(1)}} e\left(\lambda_\gamma^{(1)\,1} - \lambda_\delta^{(1)\,2}\right)$$
(2.7)

$$\prod_{n=1}^{\infty}\prod_{\alpha=1}^{M_n} E_{n1}\left(\lambda_\delta^{(1)\,2} - \lambda_\alpha^n\right) = \prod_{\gamma=1}^{M_1^{(1)}} e\left(\lambda_\delta^{(1)\,2} - \lambda_\gamma^{(1)\,1}\right) \prod_{\epsilon=1}^{M_2^{(1)}} e\left(\frac{\lambda_\delta^{(1)\,2} - \lambda_\epsilon^{(1)\,2}}{2}\right)$$

$$\prod_{\gamma=1}^{M_1^{(1)}} e\left(\frac{\lambda_\nu^{(2)\,P} - \lambda_\gamma^{(1)\,1}}{p}\right) = \prod_{(\mu,l)\neq(\nu,p)}^{M_l^{(2)}} E_{pl}(\lambda_\nu^{(2)\,P} - \lambda_\mu^{(2)\,l}) \ .$$

Here $e(x) = \frac{x+i}{x-i}$ and

$$E_{n,m}(x) = \begin{cases} e\left(\frac{x}{|n-m|}\right) e^2\left(\frac{x}{|n-m|+2}\right) \ldots e^2\left(\frac{x}{n+m-2}\right) e\left(\frac{x}{n+m}\right) & \text{if } n \neq m \\ e^2\left(\frac{x}{2}\right) e^2\left(\frac{x}{4}\right) \ldots e^2\left(\frac{x}{2n-2}\right) e\left(\frac{x}{2n}\right) & \text{if } n = m \end{cases} \tag{2.8}$$

are Takahashi's e-functions[18,19], and

(i) $M_n$ denotes the number of $\lambda$-strings of lenght $n$
(ii) $M_1^{(1)}$ denotes the number of real $\lambda^{(1)}$'s
(iii) $M_2^{(1)}$ denotes the number of $\lambda^{(1)} - \lambda^{(2)}$-strings
(iv) $M_p^{(2)}$ denotes the number of $\lambda^{(2)}$-strings of lenght $p$.

Furthermore we define for later convenience

$$M := \sum_{n=1}^{\infty} n M_n = N_\downarrow + N_h + N_l \quad ,$$

$$M^{(1)} := M_1^{(1)} + 2 M_2^{(1)} = N_h + N_l \quad ,$$

$$M^{(2)} := M_2^{(1)} + \sum_{p=1}^{\infty} p M_p^{(2)} = N_l \quad .$$

Taking the logarithm of equations (2.7) we arrive at

$$L\theta\left(\frac{\lambda_\alpha^n}{n}\right) = 2\pi I_\alpha^n + \sum_{(\beta,m) \neq (\alpha,n)} \theta_{m,n}(\lambda_\alpha^n - \lambda_\beta^m) + \sum_{\gamma=1}^{M_1^{(1)}} \theta\left(\frac{\lambda_\gamma^{(1)1} - \lambda_\alpha^n}{n}\right) + \sum_{\delta=1}^{M_2^{(1)}} \theta_{n,1}(\lambda_\delta^{(1)2} - \lambda_\alpha^n)$$

$$\sum_{n=1}^{\infty} \sum_{\alpha=1}^{M_n} \theta\left(\frac{\lambda_\gamma^{(1)1} - \lambda_\alpha^n}{n}\right) = 2\pi J_\gamma + \sum_{p=1}^{\infty} \sum_{\nu=1}^{M_p^{(2)}} \theta\left(\frac{\lambda_\gamma^{(1)1} - \lambda_\nu^{(2)p}}{p}\right) + \sum_{\delta=1}^{M_2^{(1)}} \theta(\lambda_\gamma^{(1)1} - \lambda_\delta^{(1)2})$$

$$\sum_{n=1}^{\infty} \sum_{\alpha=1}^{M_n} \theta_{n,1}(\lambda_\delta^{(1)2} - \lambda_\alpha^n) = 2\pi K_\delta + \sum_{\gamma=1}^{M_1^{(1)}} \theta(\lambda_\delta^{(1)2} - \lambda_\gamma^{(1)1}) + \sum_{\epsilon=1}^{M_2^{(1)}} \theta\left(\frac{\lambda_\delta^{(1)2} - \lambda_\epsilon^{(1)2}}{2}\right)$$

$$\sum_{\gamma=1}^{M_1^{(1)}} \theta\left(\frac{\lambda_\nu^{(2)p} - \lambda_\gamma^{(1)1}}{p}\right) = 2\pi N_\nu^p + \sum_{(\mu,l) \neq (\nu,p)} \theta_{p,l}(\lambda_\nu^{(2)p} - \lambda_\mu^{(2)l}) \quad .$$

(2.9)

Here $I_\alpha^n$, $J_\gamma$, $K_\delta$, $K_\nu^p$ are integer or half odd integer numbers, $\theta(x) = 2\arctan(x)$, and

$$\theta_{n,m}(x) = \begin{cases} \theta\left(\frac{x}{|n-m|}\right) + 2\,\theta\left(\frac{x}{|n-m|+2}\right) + \ldots + 2\,\theta\left(\frac{x}{n+m-2}\right) + \theta\left(\frac{x}{n+m}\right) & \text{if } n \neq m \\ 2\,\theta\left(\frac{x}{2}\right) + 2\,\theta\left(\frac{x}{4}\right) + \ldots + 2\,\theta\left(\frac{x}{2n-2}\right) + \theta\left(\frac{x}{2n}\right) & \text{if } n = m \end{cases} \quad . \tag{2.10}$$

The integers (or half-odd integers) are determined by the choice of cuts in the logarithms and have the following ranges

$$\left.\begin{array}{c} I_\alpha^n \\ J_\gamma \\ K_\delta \\ N_\nu^p \end{array}\right\} \left\{\begin{array}{c} \text{integer} \\ \text{half-odd int.} \end{array}\right\} \text{ if } \left\{\begin{array}{c} L + M_1^{(1)} - M_2^{(1)}\delta_{n1} + M_n - 1 \\ \sum_{n=1}^\infty M_n - \sum_{p=1}^\infty M_p^{(2)} - M_2^{(1)} \\ M_1 - M_1^{(1)} - M_2^{(1)} - 1 \\ M_1^{(1)} + M_p^{(2)} - 1 \end{array}\right\} \left\{\begin{array}{c} \text{even} \\ \text{odd} \end{array}\right. . \quad (2.11)$$

$$|I_\alpha^n| \leq \frac{1}{2} \left( L + M_1^{(1)} + M_2^{(1)}(2 - \delta_{n1}) - \sum_{m=1}^\infty t_{mn} M_m - 1 \right) ,$$

$$|J_\gamma| \leq \frac{1}{2} \left( \sum_{n=1}^\infty M_n - \sum_{p=1}^\infty M_p^{(2)} - M_2^{(1)} \right) - 1 ,$$

$$(2.12)$$

$$|K_\delta| \leq \frac{1}{2} \left( M_1 + 2 \sum_{m=2}^\infty M_m - M_1^{(1)} - M_2^{(1)} - 1 \right) ,$$

$$|N_\nu^p| \leq \frac{1}{2} \left( M_1^{(1)} - \sum_{q=1}^\infty t_{pq} M_q^{(2)} - 1 \right) .$$

Here $t_{mn} = 2 \min\{m,n\} - \delta_{m,n}$. These equations will be needed to determine the ground state and the excitation spectrum at zero temperature.

Insertion of (2.3)-(2.6) into (2.2) leads to the follwoing expressions for energy and momentum in terms of the (real) centers of the strings

$$E(\{\lambda_\alpha^n\}) = L(1 + h - \mu - \frac{U}{4}) - \sum_{n=1}^\infty \sum_{\alpha=1}^{M_n} \left( \frac{4n}{(\lambda_\alpha^n)^2 + n^2} + 2hn \right)$$

$$+ M_1^{(1)}(\mu + \frac{U}{2} + h) + M_2^{(1)}(U + 2h) - \sum_{p=1}^\infty 2\mu p \, M_p^{(2)} , \quad (2.13)$$

$$P(\{\lambda_\alpha^n\}) = (L+1)\pi + i \sum_{n=1}^\infty \sum_{\alpha=1}^{M_n} \ln(e\left(\frac{\lambda_\alpha^n}{n}\right)) .$$

Inserting the logarithmic form of the Bethe equations (2.9) into our expression for the momentum (2.13), it is possible to express the momentum completely in terms of the integers present in (2.9)

$$P = \sum_{n=1}^{\infty} \sum_{\alpha=1}^{M_n} \theta\left(\frac{\lambda_\alpha^n}{n}\right) - \pi$$

$$= \frac{2\pi}{L} \left( \sum_{n=1}^{\infty} \sum_{\alpha=1}^{M_n} I_\alpha^n + \sum_{\gamma=1}^{M_1^{(1)}} J_\gamma + \sum_{\delta=1}^{M_2^{(1)}} K_\delta - \sum_{p=1}^{\infty} \sum_{\nu=1}^{M_p^{(2)}} N_\nu^p \right) + \pi\left(L + 1 - \sum_{n=1}^{\infty} M_n\right) .$$
(2.14)

## 3. Thermodynamic Bethe Ansatz

In this section we will derive an infinite set of coupled integral equations (TBA equations) that determine the thermodynamical properties of the model in terms of the values of the magnetic field, chemical potential and coupling $U$. We will follow very closely the route taken by Takahashi in [18,19][4], which is a generalization of Yang and Yang's method[21] to the case of an infinite number of string solutions of the Bethe equations and the nested Bethe Ansatz. Before we start with the derivations we recall that it is sufficient to consider values of the chemical potential $\mu \leq 0$. One might expect problems with the formalism at zero chemical potential for the following reason: It was shown in [7] that for $\mu = 0$ the compressibility becomes infinite, and the ground state structure is governed by the $u(2|2)$ symmetry, which is a structure that complements the Bethe Ansatz rather than being encompassed by it. However, our analysis at zero temperature shows, that taking the limit $\mu \to 0$ of the TBA equations derived below (from the region $\mu < 0$) gives the correct results. This leads us to believe that the TBA equations also give the correct thermodynamics at $\mu = 0$ for $T > 0$. To derive the TBA equations we first rewrite the Bethe Ansatz equations (2.9) using the concept of counting functions[21]. Defining $z_n(\lambda) = 2\pi \frac{I_\alpha^n}{L}$ (and similarly $y_1(\lambda^{(1)}) = 2\pi \frac{J_\gamma}{L}$, etc.) and then taking the thermodynamic limit $L \to \infty$ in which we replace sums by integrals weighted by the density functions we arrive at

$$z_n(\lambda) = \theta\left(\frac{\lambda}{n}\right) - \sum_{m=1}^{\infty} \int_{-\infty}^{+\infty} d\Lambda \, \theta_{m,n}(\lambda - \Lambda) \rho_m(\Lambda)$$
$$- \int_{-\infty}^{+\infty} d\Lambda \, \theta\left(\frac{\Lambda - \lambda}{n}\right) \rho_1^{(1)}(\Lambda) - \int_{-\infty}^{+\infty} d\Lambda \, \theta_{n,1}(\Lambda - \lambda) \rho_2^{(1)}(\Lambda) ,$$

$$y_1(\lambda^{(1)}) = \sum_{n=1}^{\infty} \int_{-\infty}^{+\infty} d\Lambda \, \theta\left(\frac{\lambda^{(1)} - \Lambda}{n}\right) \rho_n(\Lambda)$$
(3.1)
$$- \sum_{p=1}^{\infty} \int_{-\infty}^{+\infty} d\Lambda \, \theta\left(\frac{\lambda^{(1)} - \Lambda}{p}\right) \rho_p^{(2)}(\Lambda) - \int_{-\infty}^{+\infty} d\Lambda \, \theta(\lambda^{(1)} - \Lambda) \rho_2^{(1)}(\Lambda) ,$$

---
[4]For a review see [20].

$$y_2(\lambda^{(1)}) = \sum_{n=1}^{\infty} \int_{-\infty}^{+\infty} d\Lambda \; \theta_{n,1}(\lambda^{(1)}-\Lambda)\rho_n(\Lambda) -$$

$$- \int_{-\infty}^{+\infty} d\Lambda \; \theta(\lambda^{(1)}-\Lambda)\rho_1^{(1)}(\Lambda) - \int_{-\infty}^{+\infty} d\Lambda \; \theta(\tfrac{\lambda^{(1)}-\Lambda}{2})\rho_2^{(1)}(\Lambda) \quad , \qquad (3.2)$$

$$x_p(\lambda^{(2)}) = \int_{-\infty}^{+\infty} d\Lambda \; \theta(\tfrac{\lambda^{(2)}-\Lambda}{p})\rho_1^{(1)}(\Lambda) - \sum_{q=1}^{\infty} \int_{-\infty}^{+\infty} d\Lambda \; \theta_{p,q}(\lambda^{(2)}-\Lambda)\rho_q^{(2)}(\Lambda) \quad .$$

Note that in the above equations all spectral parameters are allowed to go to infinity, which corresponds to the action of the symmetry algebra $u(2|2)$. Thus the Thermodynamic Bethe Ansatz takes into account not only Bethe states (which are only the lowest weight states of $u(2|2)$ multiplets) but also states obtained from those by acting with $u(2|2)$ raising operators. This is absolutely essential because *all* states in the Hilbert space contribute to the thermodynamic properties and have to be taken into account.

The densities for the vacancies (allowed integer numbers parametrising solutions of the Bethe equations in the sense of (2.9)) are given as usual as the derivaties of the counting functions with respect to the spectral parameters

$$\frac{dz_n(\lambda)}{d\lambda} = 2\pi \left( \rho_n(\lambda) + \bar{\rho}_n(\lambda) \right) \quad ,$$

$$\frac{dy_1(\lambda^{(1)})}{d\lambda^{(1)}} = 2\pi \left( \rho_1^{(1)}(\lambda^{(1)}) + \bar{\rho}_1^{(1)}(\lambda^{(1)}) \right) \quad ,$$

$$\frac{dy_2(\lambda^{(1)})}{d\lambda^{(1)}} = 2\pi \left( \rho_2^{(1)}(\lambda^{(1)}) + \bar{\rho}_2^{(1)}(\lambda^{(1)}) \right) \quad , \qquad (3.3)$$

$$\frac{dx_p(\lambda^{(2)})}{d\lambda^{(2)}} = 2\pi \left( \rho_p^{(2)}(\lambda^{(2)}) + \bar{\rho}_p^{(2)}(\lambda^{(2)}) \right) \quad .$$

Here the $\rho$'s are the distribution functions of spectral parameters corresponding to the taken integers ("particles") whereas the $\bar{\rho}$'s are densities related to the allowed integers which are not occupied ("holes"). Defining the integral operators

$$A_{n,m} * f \Big|_\lambda = \delta_{nm} f(\lambda) + \frac{1}{2\pi} \frac{d}{d\lambda} \int_{-\infty}^{\infty} d\lambda' \Theta_{n,m}(\lambda-\lambda') \, f(\lambda') \quad ,$$

$$[n] * f \Big|_\lambda = \frac{1}{2\pi} \int_{-\infty}^{\infty} d\lambda' \frac{2n}{n^2+(\lambda-\lambda')^2} \, f(\lambda') \qquad (3.4)$$

and differentiating (3.1),(3.2) with respect to the spectral parameters the Bethe equations (in the thermodynamic limit) turn into

$$\bar{\rho}_n(\lambda) = \frac{1}{2\pi}\frac{2n}{n^2+\lambda^2} - \sum_{m=1}^{\infty} A_{n,m} * \rho_m \bigg|_\lambda + (A_{n,1} - \delta_{n,1}) * \rho_2^{(1)} \bigg|_\lambda + [n] * \rho_1^{(1)} \bigg|_\lambda ,$$

$$\bar{\rho}_1^{(1)}(\lambda^{(1)}) = -\rho_1^{(1)}(\lambda^{(1)}) + \sum_{n=1}^{\infty}[n] * \rho_n \bigg|_{\lambda^{(1)}} - [1] * \rho_2^{(1)} \bigg|_{\lambda^{(1)}} - \sum_{p=1}^{\infty}[p] * \rho_p^{(2)} \bigg|_{\lambda^{(1)}} ,$$

$$\bar{\rho}_2^{(1)}(\lambda^{(1)}) = -\rho_2^{(1)}(\lambda^{(1)}) + \sum_{n=1}^{\infty}(A_{n,1} - \delta_{n,1}) * \rho_n \bigg|_{\lambda^{(1)}} - [1] * \rho_1^{(1)} \bigg|_{\lambda^{(1)}} - [2] * \rho_2^{(1)} \bigg|_{\lambda^{(1)}} ,$$

$$\bar{\rho}_p^{(2)}(\lambda^{(2)}) = -\sum_{p=1}^{\infty} A_{p,q} * \rho_q^{(2)} \bigg|_{\lambda^{(2)}} + [p] * \rho_1^{(1)} \bigg|_{\lambda^{(2)}} .$$

(3.5)

This set of equations expresses the densities for the "holes" in terms of the densities for the "particles". The set of equations determining the densities $\rho$ are obtained from the condition that the system be in thermodynamic equilibrium, *i.e.* from minimization of the thermodynamic potential

$$\frac{F}{L} = \frac{1}{L}(E - TS) = 1 + h - \mu - \frac{U}{4}$$
$$-\sum_{n=1}^{\infty}\int_{-\infty}^{\infty} d\lambda \left\{ \left(\tfrac{4n}{\lambda^2+n^2}+2nh\right) \rho_n + T\left((\rho_n + \bar{\rho}_n)\ln(\rho_n + \bar{\rho}_n) - \rho_n \ln(\rho_n) - \bar{\rho}_n \ln(\bar{\rho}_n)\right) \right\}$$
$$+\int_{-\infty}^{\infty} d\lambda^{(1)} \left\{ (\mu+\tfrac{U}{2}+h)\, \rho_1^{(1)} - T\left((\rho_1^{(1)} + \bar{\rho}_1^{(1)})\ln(\rho_1^{(1)} + \bar{\rho}_1^{(1)}) - \rho_1^{(1)} \ln(\rho_1^{(1)}) - \bar{\rho}_1^{(1)} \ln(\bar{\rho}_1^{(1)})\right) \right\}$$
$$+\int_{-\infty}^{\infty} d\lambda^{(1)} \left\{ (U+2h)\, \rho_2^{(1)} - T\left((\rho_2^{(1)} + \bar{\rho}_2^{(1)})\ln(\rho_2^{(1)} + \bar{\rho}_2^{(1)}) - \rho_2^{(1)} \ln(\rho_2^{(1)}) - \bar{\rho}_2^{(1)} \ln(\bar{\rho}_2^{(1)})\right) \right\}$$
$$-\sum_{p=1}^{\infty}\int_{-\infty}^{\infty} d\lambda^{(2)} \left\{ 2\mu p\, \rho_p^{(2)} + T\left((\rho_p^{(2)} + \bar{\rho}_p^{(2)})\ln(\rho_p^{(2)} + \bar{\rho}_p^{(2)}) - \rho_p^{(2)} \ln(\rho_p^{(2)}) - \bar{\rho}_p^{(2)} \ln(\bar{\rho}_p^{(2)})\right) \right\} .$$

(3.6)

At this point it is convenient to introduce the quantities

$$\alpha_n = \frac{\bar{\rho}_n}{\rho_n},\ \beta_1 = \frac{\bar{\rho}_1^{(1)}}{\rho_1^{(1)}},\ \beta_2 = \frac{\bar{\rho}_2^{(1)}}{\rho_2^{(1)}},\ \gamma_p = \frac{\bar{\rho}_p^{(2)}}{\rho_p^{(2)}} . \tag{3.7}$$

Varying the potential $F$ under the constraints (3.5) we obtain the following conditions for thermodynamic equilibrium

$$\ln(1 + \alpha_n(\lambda)) = -\frac{1}{T}\left(\frac{4n}{\lambda^2 + n^2} + 2nh\right) + \sum_{m=1}^{\infty} A_{m,n} * \ln\left(1 + \frac{1}{\alpha_m}\right)\bigg|_\lambda$$
$$- [n] * \ln\left(1 + \frac{1}{\beta_1}\right)\bigg|_\lambda - (A_{n,1} - \delta_{n,1}) * \ln\left(1 + \frac{1}{\beta_2}\right)\bigg|_\lambda ,$$

(3.8)

$$\ln(\beta_1(\lambda^{(1)})) = \frac{\mu + \frac{U}{2} + h}{T} - \sum_{n=1}^{\infty}[n] * \ln\left(1 + \frac{1}{\alpha_n}\right)\bigg|_{\lambda^{(1)}}$$
$$+ [1] * \ln\left(1 + \frac{1}{\beta_2}\right)\bigg|_{\lambda^{(1)}} - \sum_{p=1}^{\infty}[p] * \ln\left(1 + \frac{1}{\gamma_p}\right)\bigg|_{\lambda^{(1)}} ,\quad (3.9)$$

$$\ln(\beta_2(\lambda^{(1)})) = \frac{U + 2h}{T} - \sum_{n=1}^{\infty}(A_{n,1} - \delta_{n,1}) * \ln\left(1 + \frac{1}{\alpha_n}\right)\bigg|_{\lambda^{(1)}}$$
$$+ [2] * \ln\left(1 + \frac{1}{\beta_2}\right)\bigg|_{\lambda^{(1)}} + [1] * \ln\left(1 + \frac{1}{\beta_1}\right)\bigg|_{\lambda^{(1)}} ,\quad (3.10)$$

$$\ln(1 + \gamma_p(\lambda^{(2)})) = -\frac{2\mu p}{T} + \sum_{q=1}^{\infty} A_{q,p} * \ln\left(1 + \frac{1}{\gamma_q}\right)\bigg|_{\lambda^{(2)}} + [p] * \ln\left(1 + \frac{1}{\beta_1}\right)\bigg|_{\lambda^{(2)}} ,\quad (3.11)$$

The thermodynamic potential at equilibrium takes the form

$$\frac{F}{L} = 1 + h - \mu - \frac{U}{4} - T\sum_{n=1}^{\infty}\frac{1}{2\pi}\int_{-\infty}^{\infty}d\lambda\ \ln\left(1 + \frac{1}{\alpha_n(\lambda)}\right)\frac{2n}{n^2 + \lambda^2} . \quad (3.12)$$

The TBA equations (3.8)-(3.12) determine all thermodynamic properties at finite temperature as functions of the chemical potential $\mu$, the magnetic field $h$ and the coupling $U$.

Using methods similar to the ones employed in [22] it is possible to recast (3.8) and (3.11) in "tridiagonal form"

$$\ln(\alpha_n(\lambda)) = \hat{p} * \ln(1 + \alpha_{n-1})\bigg|_{\lambda} + \hat{p} * \ln(1 + \alpha_{n+1})\bigg|_{\lambda}$$
$$- \left(\frac{4\pi}{4T\ cosh(\frac{\lambda\pi}{2})} + \hat{p} * \ln\left(1 + \frac{1}{\beta_1}\right)\bigg|_{\lambda}\right)\delta_{n,1} - \hat{p} * \ln\left(1 + \frac{1}{\beta_2}\right)\bigg|_{\lambda}\delta_{n,2} ,$$
$$\ln(\gamma_s(\lambda^{(2)})) = \hat{p} * \ln(1 + \gamma_{s-1})\bigg|_{\lambda^{(2)}} + \hat{p} * \ln(1 + \gamma_{s+1})\bigg|_{\lambda^{(2)}} + \hat{p} * \ln\left(1 + \frac{1}{\beta_1}\right)\bigg|_{\lambda^{(2)}}\delta_{s,1} .$$
$$(3.13)$$

Here $\hat{p}$ is an integral operator defined by

$$\hat{p} * f\bigg|_{\lambda} = \int_{-\infty}^{\infty} d\lambda' \frac{1}{4\cosh\left((\lambda - \lambda')\frac{\pi}{2}\right)} f(\lambda') .$$

We have not succeeded in simplifying (3.9) and (3.10) in a similar manner, which makes it nontrivial to take the $T \to 0$ limit of the TBA equations (see below). Similar problems have also been encountered for other models[23]. Equations (3.8)-(3.11) can be used to compute the low-temperature specific heat[24].

### The limit $T \to 0$

In order to study the zero temperature limit we first define the dressed energies according to

$$\alpha_n = \exp(\frac{\epsilon_n}{T}), \ \beta_1 = \exp(\frac{\epsilon_1^{(1)}}{T}), \ \beta_2 = \exp(\frac{\epsilon_2^{(1)}}{T}), \ \gamma_p = \exp(\frac{\epsilon_p^{(2)}}{T}) \ . \tag{3.14}$$

We will show in our discussion of the excitation spectrum in the $T = 0$ formalism below that the $\epsilon$'s really play the role of dressed energies at zero temperature. It can be shown that this interpretation also holds at finite temperature[21,25].

From (3.13) it is clear that all $\epsilon_s^{(2)}$ and all $\epsilon_n$, $n \geq 3$ are non-negative and will drop out of the r.h.s. in the $T \to 0$ limit of the TBA equations (3.8)-(3.11). The only energies that can change sign are $\epsilon_1$, $\epsilon_2$, $\epsilon_1^{(1)}$ and $\epsilon_2^{(1)}$. Investigation of the TBA equations shows that either both $\epsilon_2^{(1)}$ and $\epsilon_2$ change sign, or both of them are non-negative for all values of spectral parameters. By a completely independent computation in the $T = 0$ formalism we can show that for $T = 0$ both are nonnegative for all values of spectral parameters. In the light of this we conjecture that even for small positive temperature only $\epsilon_1$ and $\epsilon_1^{(1)}$ change signs as functions of their respective spectral parameters. The ground state in the zero temperature limit is determined by using the knowledge which of the dressed energies are non-negative to simplify (3.5). We find that in the limit $T \to 0$ only the densities $\rho_1(\lambda)$ and $\rho_1^{(1)}(\lambda^{(1)})$ are nonvanishing. It can be seen that $\rho_1(\lambda)$ is positive for values of $\lambda$ in the interval $[-A, A]$ ( where $A$ is some positive real number), and vanishes outside this interval. A similar result holds for $\rho_1^{(1)}(\lambda^{(1)})$ and $[-B, B]$. The values of $A$ and $B$ depend on the values of the chemical potential $\mu$, the magnetic field $h$ and the coupling $U$. In the limit of vanishing external magnetic field ($h \to 0$) $A$ tends to infinity, which is the case we will restrict ourselves to for most of the remainder of this paper. Using (3.5) we thus find that the densities describing the ground state for $\mu < 0$ are given as the solutions of the following coupled integral equations

$$\rho_1(\lambda) = \frac{1}{2\pi} \frac{2}{1+\lambda^2} - \frac{1}{2\pi} \int_{-\infty}^{\infty} d\lambda' \frac{4}{4+(\lambda-\lambda')^2} \rho_1(\lambda')$$
$$+ \frac{1}{2\pi} \int_{-B}^{+B} d\lambda^{(1)} \frac{2}{1+(\lambda-\lambda^{(1)})^2} \rho_1^{(1)}(\lambda^{(1)}) \ , \tag{3.15}$$
$$\rho_1^{(1)}(\lambda^{(1)}) = \frac{1}{2\pi} \int_{-\infty}^{\infty} d\lambda \frac{2}{1+(\lambda-\lambda^{(1)})^2} \rho_1(\lambda) \ .$$

The value of $B$ is determined by the requirement that

$$\int_{-B}^{B} d\lambda^{(1)} \rho_1^{(1)}(\lambda^{(1)}) = \frac{N_h}{L} = 1 - \frac{N_e}{L} = 1 - D . \tag{3.16}$$

In the grand canonical ensemble $B$ is given as a function of the chemical potential and the coupling $U$ via the boundary condition $\epsilon_1^{(1)}(\pm B) = 0$. This enables us to solve for $D$ as a function of $\mu$.

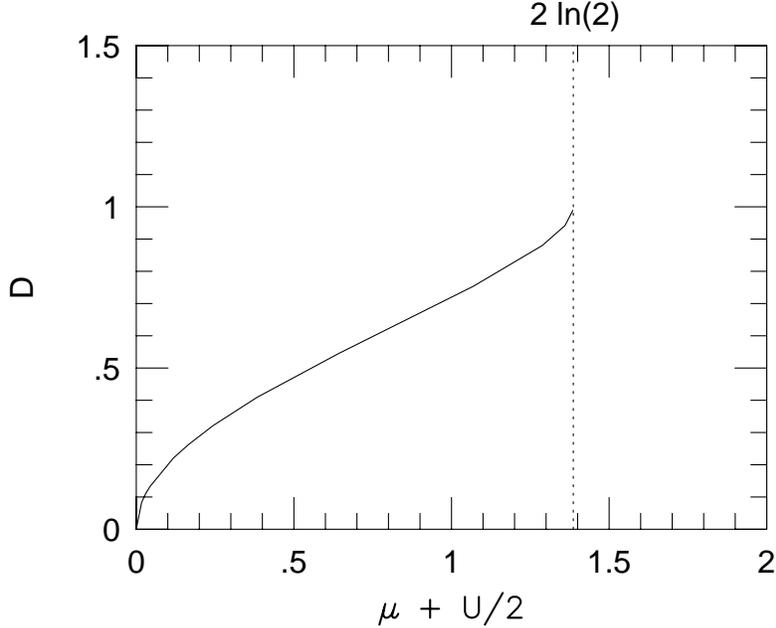

Figure 1: $T = 0$ Ground state density as a function of chemical potential and coupling $U$ for $\mu < 0$

Equations (3.15) are recognized to be identical to the ground state integral equations of the supersymmetric $t$-$J$ model (in zero external magnetic field) in the Sutherland representation[6,26,27]. An investigation of the dependence of the integration boundary $B$ on the chemical potential $\mu$ and coupling $U$ shows that the ground state of the $u(2|2)$ model for $\mu < 0$ is actually identical to the ground state of the supersymmetric $t$-$J$ model for a value of the $t$-$J$ chemical potential $\mu_{tJ} = \mu + \frac{U}{2}$. This confirms a result stated in [7] for the $u(2|2)$ model in any dimension. From Figure 1 we see that the ground state is equal to the empty vacuum $|0\rangle$ in the region where $\mu + \frac{U}{2} \leq 0$. For $\mu + \frac{U}{2} \geq 2\ln(2)$ the ground state is equal to the half-filled ground state of the supersymmetric $t$-$J$ model. Thus we exactly reproduce the $\mu < 0$ part of the zero temperature phase diagram derived in [10], which is pictured in Figure 2.

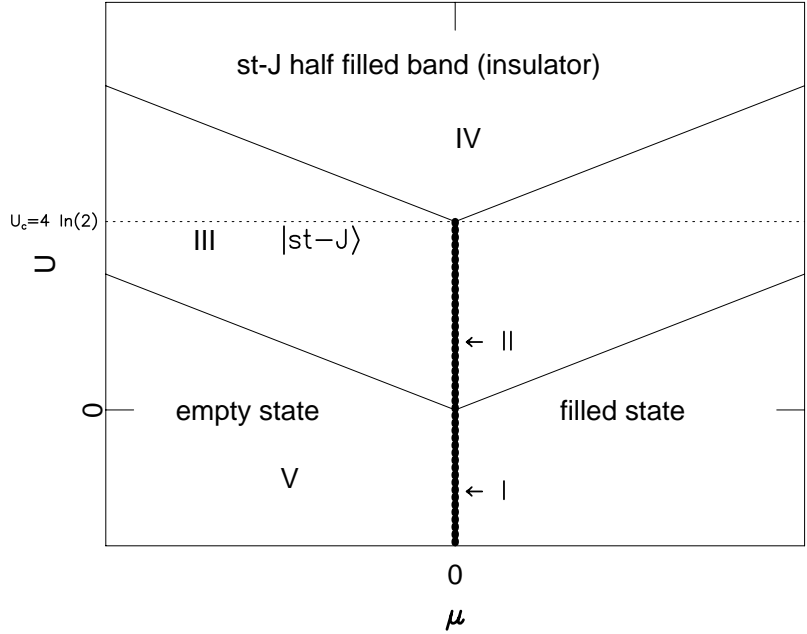

Figure 2 : Ground states in the grand canonical ensemble

Taking now the $T \to 0$ limit in (3.8)-(3.11) and using (3.14) we arrive at the following equations for the dressed energies at zero temperature and $h = 0$

$$\begin{aligned}
\epsilon_n(\lambda) &= -\frac{4n}{n^2 + \lambda^2} - \frac{1}{2\pi} \int_{-\infty}^{\infty} d\lambda' \frac{d}{d\lambda} \theta_{1n}(\lambda - \lambda') \, \epsilon_1(\lambda') \\
&\quad + \frac{1}{2\pi} \int_{-B}^{+B} d\lambda^{(1)} \, \frac{2n}{n^2 + (\lambda^{(1)} - \lambda)^2} \, \epsilon_1^{(1)}(\lambda^{(1)}) \;, \\
\epsilon_1^{(1)}(\lambda^{(1)}) &= \mu + \frac{U}{2} + \frac{1}{2\pi} \int_{-\infty}^{\infty} d\lambda \frac{2}{1 + (\lambda - \lambda^{(1)})^2} \, \epsilon_1(\lambda) \;, \\
\epsilon_2^{(1)}(\lambda^{(1)}) &= U + \frac{1}{2\pi} \int_{-\infty}^{\infty} d\lambda \frac{4}{4 + (\lambda - \lambda^{(1)})^2} \, \epsilon_1(\lambda) \\
&\quad - \frac{1}{2\pi} \int_{-B}^{+B} d\Lambda \frac{2}{1 + (\lambda^{(1)} - \Lambda)^2} \, \epsilon_1^{(1)}(\Lambda) \;, \\
\epsilon_p^{(2)}(\lambda^{(2)}) &= -2p\mu - \frac{1}{2\pi} \int_{-B}^{+B} d\Lambda \frac{2p}{p^2 + (\lambda^{(2)} - \Lambda)^2} \, \epsilon_1^{(1)}(\Lambda)
\end{aligned} \quad (3.17)$$

In section 4 below we will reobtain these equations as describing the excitation energies of the low-lying excitations above the ground state in the zero temperature formalism. The integral equations for the dressed energies cannot be solved analytically for arbitrary filling. However one can decouple them by Fourier transforming and finds

$$\epsilon_1(\lambda) = -\frac{2\pi}{2\cosh(\frac{\lambda\pi}{2})} - \frac{1}{8\pi}\int_{-B}^{B} d\Lambda \int_{-\infty}^{\infty} d\mu \; \frac{2}{1+(\Lambda-\mu)^2} \frac{1}{\cosh(\frac{[\lambda-\Lambda]\pi}{2})} \; \epsilon_1(\mu)$$

$$+ (\mu + \frac{U}{2})\int_{-B}^{B} d\Lambda \; \frac{1}{4\cosh(\frac{[\lambda-\Lambda]\pi}{2})} \; , \quad (3.18)$$

$$\epsilon_1^{(1)}(\Lambda) = \mu + \frac{U}{2} - 4\pi \; R(\Lambda) + \int_{-B}^{B} d\Lambda' \; R(\Lambda-\Lambda') \; \epsilon_1^{(1)}(\Lambda') \; ,$$

where $R(x) = \frac{1}{4\pi}\int_{-\infty}^{\infty} dt \; \frac{1}{1+(x-t)^2} \frac{1}{\cosh(\frac{\pi}{2}t)}$.

### The limit $T \to \infty$

We now turn to the study of the infinite temperature limit. The analysis is a straightforward generalization of Takahashi's method for the Fermi gas with $\delta$-function interaction[28]. In this section we will keep the possibility of having a finite magnetic field $h$. The infinite temperature limit is of interest because it gives an estimate for the number of Bethe states in the thermodynamic limit, thus providing a rough check on whether the Bethe states form a basis of the complete Hilbert space. Furthermore it is possible to obtain a high temperature expansion for the Gibbs free energy in this limit.

We first observe that for $T \to \infty$ all driving terms tend to zero in the integral equations and $\alpha_n$, $\beta_1$, $\beta_2$, $\gamma_s$ all turn into constants. Therefore the integral equations turn into coupled algebraic equations, which can be solved analytically. We first note that in the infinite temperature limit insertion of (3.8) for $n = 1$ into (3.10) leads to the following simple relation, which allows us to eliminate $\beta_2$ from all other equations, which we will do below.

$$\beta_2 = \frac{1}{\alpha_1} e^{\frac{U}{T}} \longrightarrow \frac{1}{\alpha_1} \; .$$

The infinite system of coupled integral equations (3.13) turns into

$$\begin{aligned}
(\alpha_1)^2 &= \frac{(1+\alpha_2)}{(1+\frac{1}{\beta_1})} \\
(\alpha_2)^2 &= \frac{(1+\alpha_1)(1+\alpha_3)}{(1+\frac{1}{\beta_2})} \\
(\alpha_n)^2 &= (1+\alpha_{n-1})(1+\alpha_{n+1}), \; n \geq 3.
\end{aligned} \quad , \quad \begin{aligned}
(\gamma_1)^2 &= (1+\gamma_2)(1+\frac{1}{\beta_1}) \\
(\gamma_s)^2 &= (1+\gamma_{s-1})(1+\gamma_{s+1}), \; s \geq 2 \; ,
\end{aligned}$$
$$(3.19)$$

and we still have to consider the equations for $\alpha_1$, $\beta_1$, and $\gamma_1$

$$\ln(1+\alpha_1) = 2\sum_{n=1}^{\infty} \ln\left(1+\frac{1}{\alpha_n}\right) - \ln\left(1+\frac{1}{\beta_1}\right) - \ln\left(1+\frac{1}{\beta_2}\right) - \frac{2h}{T},$$

$$\ln(\beta_1) = \ln\left(1+\frac{1}{\beta_2}\right) - \sum_{n=1}^{\infty}\ln\left(1+\frac{1}{\alpha_n}\right) - \sum_{p=1}^{\infty}\ln\left(1+\frac{1}{\gamma_p}\right) + \frac{\mu+h}{T} \quad (3.20)$$

$$\ln(1+\gamma_1) = \ln\left(1+\frac{1}{\beta_1}\right) + 2\sum_{p=1}^{\infty}\ln\left(1+\frac{1}{\gamma_p}\right) - \frac{2\mu}{T}.$$

Using (3.19) in (3.20) we find the "boundary conditions"

$$\lim_{N\to\infty}\ln\left(\frac{\alpha_N+1}{\alpha_{N+1}+1}\right) = \frac{2h}{T}, \quad \ln(\beta_1) = -\frac{1}{2}\ln(1+\gamma_1),$$

$$\lim_{N\to\infty}\ln\left(\frac{\gamma_N+1}{\gamma_{N+1}+1}\right) = \frac{2\mu}{T}. \quad (3.21)$$

The recursion relations (3.19) have the solutions

$$\gamma_s = (f(s))^2 - 1, \quad \beta_1 = \frac{1}{(f(0))^2 - 1},$$

$$\alpha_n = (g(n))^2 - 1, n\geq 2, \quad (\alpha_1)^2 = \left(\frac{g(2)}{f(0)}\right)^2, \quad (3.22)$$

where

$$f(s) = \frac{ba^s - b^{-1}a^{-s}}{a - a^{-1}}, \quad g(n) = \frac{c^n - c^{-n}}{c - c^{-1}}.$$

Conditions (3.21) now determine $a$, $b$, and $c$ as functions of $T$, $\mu$ and $h$ (Note that like Takahashi we keep the ratios $\frac{\mu}{T}$ and $\frac{h}{T}$ finite for the time being, but take $\frac{U}{T}$ to zero.)

$$\alpha_1 = \frac{\cosh\left(\frac{h}{T}\right)}{\cosh\left(\frac{\mu}{T}\right)} = (\beta_2)^{-1},$$

$$\alpha_n = \left(\frac{\sinh\left(\frac{nh}{T}\right)}{\sinh\left(\frac{h}{T}\right)}\right)^2 - 1, \; n\geq 2,$$

$$\beta_1 = \frac{1}{4\left(\cosh\left(\frac{\mu}{T}\right)\right)^2 - 1}, \quad (3.23)$$

$$\gamma_s = \left(\frac{\sinh\left(\frac{(s+2)\mu}{T}\right)}{\sinh\left(\frac{\mu}{T}\right)}\right)^2 - 1, \; s\geq 1.$$

The thermodynamic potential $F$ can be evaluated exactly in the infinite temperature limit. Clearly the entropy contribution is dominating and the result thus gives us an estimate for the total number of states in the Hilbert space. This is because

$$\frac{F}{L} = -\frac{T}{L} \ln\left(tr(e^{\frac{-H(\mu,U,h)}{T}})\right) \xrightarrow[T \to \infty]{} -\frac{T}{L} \ln(tr(id)) = -T\ln(4) \ .$$

The formalism developed above leads to the following result

$$\frac{F}{L} = 1 + h - \mu - \frac{U}{4} - T\sum_{n=1}^{\infty} \ln\left(1 + \frac{1}{\alpha_n}\right) \to -T\ln\left(2\ cosh\left(\frac{\mu}{T}\right) + 2\ cosh\left(\frac{h}{T}\right)\right) \ .$$

This obviously approaches $-T\ln(4)$ and we thus conclude that the dimension of the Hilbert space is roughly $4^L$ (because there can be corrections of order $\ln(L)$ which we cannot determine using the techniques employed above), which is the correct result. A rigorous clarification of the question whether the set of Bethe states is complete is rather difficult, as counting arguments as employed in [14] (for the XXX model) or [13] (for the Hubbard model) fail due to the presence of zero-norm states[29][5]. Thus it is necessary to first determine the norms of the wavefunctions[30] and then count the non-vanishing states[31].

## 4. Excitation spectrum at zero temperature

In this section we will derive the spectrum of the low-lying excitations over the ground states in the various regions in the $\mu - U$ plane at $T = 0$. The external magnetic field $h$ is taken to be zero. We will make use of the "standard" $T = 0$-formalism that goes back to Hulthén and des Cloizeaux and Pearson[32,33], the results of which are found to agree with the $T \to 0$ limit of the thermodynamics as discussed in section 3. In [10] we have proved, that the Bethe Ansatz only provides lowest weight states of the $u(2|2)$ symmetry algebra. Therefore the spectrum of the $u(2|2)$-symmetric part $H^0$ of the hamiltonian (1.1) is highly degenerate. The chemical potential and Hubbard interaction terms break most of this invariances. The only degeneracies left are due to the spin-$SU(2)$, which is not broken as the magnetic field is taken to be zero. In order to construct the low-lying excitations over the ground states (which were determined in [7], and again from the TBA in section 3 above) we will first use the usual Bethe Ansatz techniques to obtain the excitations given by the Bethe Ansatz. We will then construct the complete $u(2|2)$ multiplet of elementary excitations by acting with the $u(2|2)$ lowering operators.

---

[5]This means that the wave functions to certain solutions of the Bethe equations vanish.

### 4.1. Ground State and Excitations in Region III

Region III in Figure 2 is the region where $\mu < 0$ $0 < \mu + \frac{U}{2} < 2\ln(2)$. We have shown is section 3 that the ground state in this region is given by the ground state of the supersymmetric $t$-$J$ model for the induced value of chemical potential $\mu_{tJ} = \mu + \frac{U}{2}$. We will now proceed to construct a complete set of excitations over this ground state. Naturally we will encounter all excitations that are already present in the supersymmetric $t$-$J$ model[6]. In addition to these we will find a large number of "localonic" excitations, that is excitations involving local electron pairs. We will start with a discussion of the ground state and then give a complete account of all $t$-$J$-like excitations. The BBFF representation of the Bethe equations[10] we have chosen in this paper reduces to the Sutherland representation[6] of the supersymmetric $t$-$J$ model in the sector without local electron pairs. The ground state and excitation spectrum of the $t$-$J$ model in this representation was studied to by Bares, Blatter and Ogata in [26] (see also [34]). They studied the spin-singlet and spin-triplet excitations, and the elementary charge excitations. Förster and Karowski[12] completed this analysis by taking into account also the $u(1|2)$-descendants of the Bethe states. Below we will perform a similar analysis for the $u(2|2)$ case. In order to discuss the $u(2|2)$ structure we will need the commutation relations (1.9) between the hamiltonian (1.1) and the symmetry generators of the $u(2|2)$ algebra.

<center><em>Ground state</em></center>

The derivation of the ground state and the excitation spectrum in the zero temperature formalism is based on equations (2.9)-(2.12). In the BBFF representation and the region $\mu < 0$ the ground state is given by the Bethe Ansatz (In the BFFB representation, which reduces to the Lai repr. of the supersymmetric $t$-$J$ model in the sector without local electron pairs, the ground state is *not* given by the Bethe Ansatz. It is obtained by acting with $Q_1^\dagger Q_{-1}^\dagger$ on the Bethe state in the multiplet. This action is equivalent to sending two spectral parameters in the Bethe state to infinity, which is how the ground state in the Lai repr. was constructed by Bares *et al.*). It is the lowest weight state of a $u(2|2)$ multiplet of dimension 16. The ground state is in general degenerate *with other Bethe Ansatz states*, but there are two cases where it is not :

(i) $L$ even, $N_h$ even, and $N_\downarrow$ odd
(ii) $L$ odd, $N_h$ odd, and $N_\downarrow$ odd.

We will consider only case (i) for simplicity.

The 15 other states in the multiplet are not degenerate with the ground state for less than half filling (as the supersymmetry is broken by the chemical potential and

---
[6]This is because all eigenstates of the supersymmetric $t$-$J$ model are also eigenstates of the $u(2|2)$ model[7].

Hubbard interaction according to (1.9)), so that there is a *unique* ground state. In terms of the distributions of roots of the Bethe equations it is described by taking

$$M_1 = N_h^{GS} + N_\downarrow^{GS} \ , \ M_n = 0 \ \ \forall n \geq 2 \ , \ M_1^{(1)} = N_h^{GS} \ , \ M_2^{(1)} = 0 = M_p^{(2)} \ ,$$

and by completely filling all vacancies for $I_\alpha^1$ between $-I_{max}^1 = \frac{1}{2}(N_h^{GS} + N_\downarrow^{GS} - 1)$ and $I_{max}^1$ (symmetrically around zero), and by filling all vacancies for $J_\gamma$ symmetrically around zero up to $J_{max} = \frac{N_h^{GS}-1}{2}$.[7]

Inserting this prescription into (2.9), subtracting equations for consecutive integers, and then taking the thermodynamic limit, we arrive at the set of coupled integral equations (3.15).

The ground state energy is computed from (2.13) (taking into account that $h = 0$)

$$\frac{E_{GS}(\mu,U)}{L} = 1 + h - \mu - \frac{U}{4} - \int_{-\infty}^{\infty} d\lambda \ \frac{4}{\lambda^2 + 1} \rho_1^{GS}(\lambda) + \frac{N_h^{GS}}{L}(\mu + \frac{U}{2})$$
$$=: 1 + h - \mu - \frac{U}{4} + \int_{-\infty}^{\infty} d\lambda \ \overset{\scriptscriptstyle(0)}{\varepsilon}_1(\lambda) \ \rho_1^{GS}(\lambda) + \int_{-B_0}^{B_0} d\Lambda \ \overset{\scriptscriptstyle(0)}{\varepsilon}_2(\Lambda) \ \rho_1^{(1)\,GS}(\lambda) \equiv \frac{E_{GS}(B_0)}{L} \ ,$$
(4.1)

where $\overset{\scriptscriptstyle(0)}{\varepsilon}_1(\lambda) = -\frac{4}{\lambda^2+1}$ and $\overset{\scriptscriptstyle(0)}{\varepsilon}_2(\Lambda) = \mu + \frac{U}{2}$ are the "bare energies" of the ground state particles in the $\lambda$- and $\lambda^{(1)}$- distributions respectively. In the second line we consider the ground state energy as a function of the integration boundary $B_0$, which will be convenient for the analysis below.

### *Spin-Wave Spectrum*

The spin-wave-type excitations are very similar to spin-waves in the spin-$\frac{1}{2}$ Heisenberg XXX antiferromagnet. The correct elementary excitations for that case were identified by L.D. Faddeev and L. Takhtajan in [14,35]. We will follow their analysis very closely. We start by trying to insert one "hole" into the distribution of integers $I_\alpha^1$, i.e. choose $M_1 = N_h^{GS} + N_\downarrow^{GS} - 1$. The number $M_1^{(1)}$ of $\lambda^{(1)}$'s is kept fixed at $N_h^{GS}$. This however changes the number of vacancies for $I_\alpha^1$ to $N_h^{GS} + N_\downarrow^{GS} + 1$, so that there are really *two* holes. We denote the spectral parameters corresponding to the missing integers by $\lambda_1^h$ and $\lambda_2^h$. By a similar analysis we find that we only can make an *even* number of holes by this procedure. In the thermodynamic limit we find the following set of coupled integral equations for a state with $2N$ holes at positions $\lambda_j^h$, $j = 1 \ldots 2N$

---
[7] In what follows we denote the number of holes and down spins in the ground state by $N_h^{GS}$ and $N_\downarrow^{GS}$ respectively.

$$\rho_1(\lambda) = \frac{1}{2\pi}\frac{2}{1+\lambda^2} - \frac{1}{2\pi}\int_{-\infty}^{\infty}d\lambda'\frac{4}{4+(\lambda-\lambda')^2}\,\rho_1(\lambda')$$
$$+ \frac{1}{2\pi}\int_{-B}^{+B}d\lambda^{(1)}\frac{2}{1+(\lambda-\lambda^{(1)})^2}\,\rho_1^{(1)}(\lambda^{(1)}) - \frac{1}{L}\sum_{j=1}^{2N}\delta(\lambda-\lambda_j^h)\,, \qquad (4.2)$$
$$\rho_1^{(1)}(\lambda^{(1)}) = \frac{1}{2\pi}\int_{-\infty}^{\infty}d\lambda\frac{2}{1+(\lambda-\lambda^{(1)})^2}\,\rho_1(\lambda)\,.$$

In the next step we now introduce a matrix integral equation formalism which will be very convenient in deriving the excitation energies. We first define the integral operator $\hat{K}$

$$\hat{K}_{\alpha\beta}*f\bigg|_{\lambda} = \int_{-a_\beta}^{a_\beta}d\mu K_{\alpha\beta}(\lambda-\mu)\,f(\mu)\quad,\ a_1=\infty,\ a_2=B\,,$$
$$K_{11}(\lambda) = -\frac{4}{4+\lambda^2},\ K_{12}(\lambda) = \frac{2}{1+\lambda^2} = K_{21}(\lambda),\ K_{22}(\lambda) = 0\,. \qquad (4.3)$$

Using the vector notation

$$\varrho_\alpha = (\rho_1,\rho_1^{(1)})^T,\quad \varrho_\alpha^{(0)} = \left(\frac{1}{\pi}\frac{1}{1+\lambda^2},\,0\right)^T \qquad (4.4)$$

we rewrite (4.2) as

$$\left(\delta_{\alpha\beta} - \frac{1}{2\pi}\hat{K}_{\alpha\beta}\right)\varrho_\beta = \varrho_\alpha^{(0)} + \frac{\psi_\alpha}{L}\,, \qquad (4.5)$$

where the "order 1 corrections" $\psi_\alpha$ are given by

$$\psi_1 = -\sum_{j=1}^{2N}\delta(\lambda-\lambda_j^h)\,,\quad \psi_2 = 0\,.$$

The excitation energy $E_{ex} = E(B) - E_{GS}(B_0)$ follows from (2.13) to be

$$E_{ex} = L\int_{-\infty}^{\infty}d\lambda\,\varepsilon_1^{(0)}(\lambda)\,(\rho_1(\lambda)-\rho_1^{GS}(\lambda)) + L\int_{-B}^{B}d\Lambda\,\varepsilon_2^{(0)}(\Lambda)\,(\rho_1^{(1)}(\Lambda)-\rho_1^{(1)GS}(\Lambda))\,. \qquad (4.6)$$

The differences of densities

$$\sigma_1(\lambda) = L(\rho_1(\lambda)-\rho_1^{GS}(\lambda))\,,$$
$$\sigma_2(\Lambda) = L(\rho_1^{(1)}(\Lambda)-\rho_1^{(1)GS}(\Lambda)) \qquad (4.7)$$

obey a similar set of coupled integral equations as the densities themselves

$$\left(\delta_{\alpha\beta} - \frac{1}{2\pi}\hat{K}_{\alpha\beta}\right)\sigma_\beta = \psi_\alpha . \tag{4.8}$$

The inverse of this equation reads (assuming invertibility)

$$\sigma_\beta = \left(id - \frac{1}{2\pi}\hat{K}\right)^{-1}_{\beta\alpha} \psi_\alpha . \tag{4.9}$$

The dressed energies at zero temperature are defined according to

$$\left(\delta_{\alpha\beta} - \frac{1}{2\pi}\hat{K}_{\alpha\beta}\right)\varepsilon_\beta = \overset{(0)}{\varepsilon}_\alpha, \tag{4.10}$$

where $\overset{(0)}{\varepsilon}_\alpha$ are defined above and we impose the boundary conditions (vanishing on the dressed energies at the Fermi boundaries)

$$\varepsilon_1(\pm\infty) = 0 , \quad \varepsilon_2(\pm B) = 0 . \tag{4.11}$$

This definition of course coincides with the integral equations for the dressed energies derived from the thermodynamics if we identify $\varepsilon_1(\lambda) = \epsilon_1(\lambda)$ and $\varepsilon_2(\Lambda) = \epsilon_1^{(1)}(\Lambda)$. We now can rewrite the expression for the excitation energy using (4.3), (4.6), (4.8), (4.9) and the inverse of (4.10)

$$E_{ex} = \sum_{\beta=1}^{2}\int_{-a_\beta}^{a_\beta} d\mu_\beta \, \overset{(0)}{\varepsilon}_\beta(\mu_\beta) \, \sigma_\beta(\mu_\beta) = \sum_{\beta=1}^{2}\int_{-a_\beta}^{a_\beta} d\mu_\beta \, \overset{(0)}{\varepsilon}_\beta(\mu_\beta) \left(id - \frac{1}{2\pi}\hat{K}\right)^{-1}_{\beta\alpha} \psi_\alpha$$

$$= \sum_{\alpha=1}^{2}\int_{-a_\alpha}^{a_\alpha} d\mu_\alpha \, \varepsilon_\alpha(\mu_\alpha) \, \psi_\alpha = -\sum_{j=1}^{2N}\epsilon_1(\lambda_j^h) ,$$

where we have used the expressions for $\psi_\alpha$ in the last equality. A numerical solution of the coupled integral equations for $\epsilon_1$ and $\epsilon_1^{(1)}$ shows that both dressed energies are gapless at $T=0$. Therefore the (multi-parametric) excitations constructed above are gapless as well. For a plot of the resulting dispersion we refer to [26]. We see that there is no charge associated with this type of excitation, whereas the component of the spin in the quantization direction and the total spin are seen to be

$$S^z = N , \quad S = N.$$

The value of the total spin follows from the value of $S^z$ and the lowest weight theorem. The simplest example is the spin triplet excitation constructed in [26]. The momentum of the excited states can be evaluated by means of (2.14). For even $N$ it is simply given by

$$P = -\frac{2\pi}{L} \sum_{j=1}^{2N} I_j^h \mod 2\pi \ .$$

For odd $N$ the integers $J_\gamma$ cannot be distributed symmetrically around zero, and there are two branches of the excitation (depending on whether we fill the "integer" $\frac{N_h^{GS}}{2}$ or $\frac{-N_h^{GS}}{2}$). The momentum is

$$P = \frac{\pi}{L}(L \pm N_h^{GS}) - \frac{2\pi}{L}\sum_{j=1}^{2N} I_j^h = \pm 2k_F - \frac{2\pi}{L}\sum_{j=1}^{2N} I_j^h \mod 2\pi \ ,$$

where the Fermi momentum is defined as $k_F = \frac{\pi N_e}{2L}$[26].

It is clear that the type of excitation contructed above does not exhaust all possible spin-wave-type excitations, as for example the spin-singlets are not included. In order to construct these we consider a state with $2N - 2$ holes in the integers $I_\alpha^1$ and one $\lambda$-string of length $n$ (with integer $I_1^n$), i.e.

$$M_1 = N_h^{GS} + N_\downarrow^{GS} - N \ , \ M_n = 1 \ , \ M_m = 0 \ \forall m \notin \{1, n\} \ ,$$
$$M_1^{(1)} = N_h^{GS} \ , \ M_2^{(1)} = 0 = M_p^{(2)} \ .$$

As compared to the "pure" hole excitation above we have an additional $n$-string, which "eats up" two vacancies in the $I^1$-distribution, so that we are left with $2N - 2$ holes. The integer for the $n$-string can be computed using (2.12) and one finds that $|I_\alpha^n| \leq (N - n)$ are the only allowed values. The spectral parameter $\lambda_1^n$ describing the center of the $n$-string can be determined from (2.9)

$$L\theta\left(\frac{\lambda_1^n}{n}\right) = \frac{2\pi I_\alpha^n}{L} + \sum_{\beta=1}^{M_1} \theta_{1,n}(\lambda_1^n - \lambda_\beta^1) + \sum_{\gamma=1}^{M_1^{(1)}} \theta\left(\frac{\lambda_\gamma^{(1)1} - \lambda_1^n}{n}\right) \ .$$

In principle this equation fixes $\lambda_1^n$ in terms of the $\lambda_j^h$'s, but the precise connection is not important for our purposes here[8].

Taking the thermodynamic limit of the Bethe equations (2.9) for this type of excitation we obtain the following set of integral equations for the densities describing the distributions of rapidities

---

[8]For $n = 2$ and $N = 1$ it follows from a computation like in [26] that $\lambda_1^2 = \frac{\lambda_1^h + \lambda_2^h}{2}$.

$$\rho_1(\lambda) = \frac{1}{2\pi} \frac{2}{1+\lambda^2} - \frac{1}{2\pi} \int_{-\infty}^{\infty} d\lambda' \frac{4}{4+(\lambda-\lambda')^2} \rho_1(\lambda')$$

$$+ \frac{1}{2\pi} \int_{-B}^{+B} d\lambda^{(1)} \frac{2}{1+(\lambda-\lambda^{(1)})^2} \rho_1^{(1)}(\lambda^{(1)})$$

$$- \frac{1}{L} \sum_{j=1}^{2N-2} \delta(\lambda - \lambda_j^h) - \frac{1}{2\pi L} \left( \frac{2(n-1)}{(n-1)^2 + (\lambda-\lambda_1^n)^2} + \frac{2(n+1)}{(n+1)^2 + (\lambda-\lambda_1^n)^2} \right),$$

$$\rho_1^{(1)}(\lambda^{(1)}) = \frac{1}{2\pi} \int_{-\infty}^{\infty} d\lambda \frac{2}{1+(\lambda-\lambda^{(1)})^2} \rho_1(\lambda) + \frac{1}{2\pi L} \left( \frac{2n}{n^2 + (\lambda-\lambda_1^n)^2} \right).$$

(4.12)

Defining

$$\psi_1 = -\sum_{j=1}^{2N} \delta(\lambda - \lambda_j^h) - \frac{1}{2\pi} \left( \frac{2(n-1)}{(n-1)^2 + (\lambda-\lambda_1^n)^2} + \frac{2(n+1)}{(n+1)^2 + (\lambda-\lambda_1^n)^2} \right)$$

$$\psi_2 = \frac{1}{2\pi} \frac{2n}{n^2 + (\lambda-\lambda_1^n)^2},$$

we arrive at a matrix equation for the backflow distribution functions $\sigma_{1,2}$ (defined in (4.7)) which is of the same form as (4.8). After similar manipulations as for the case with only holes we arrive at the following expression for the excitation energy

$$E_{ex} = \overset{(o)}{\varepsilon}_n(\lambda_1^n) + \sum_{\alpha=1}^{2} \int_{-a_\alpha}^{a_\alpha} d\mu_\alpha \, \varepsilon_\alpha(\mu_\alpha) \, \psi_\alpha \, , \qquad (4.13)$$

where $\overset{(o)}{\varepsilon}_n(\lambda_1^n) = -\frac{4n}{n^2+(\lambda_1^n)^2}$ is the "bare" energy of the $n$-string, and $\varepsilon_\alpha$ are defined by (4.10). Recall that $\varepsilon_{1,2}$ is a matrix notation for $\epsilon_1$ and $\epsilon_1^{(1)}$.

This expression breaks up into contributions from the holes and the $n$-string. Defining the dressed energy $\epsilon_n(\lambda_1^n)$ of the $n$-string according to

$$\epsilon_n(\lambda) = -\frac{4}{n^2+\lambda^2} - \frac{1}{2\pi} \int_{-\infty}^{\infty} d\lambda' \frac{d}{d\lambda} \theta_{1n}(\lambda - \lambda') \, \epsilon_1(\lambda')$$

$$+ \frac{1}{2\pi} \int_{-B}^{+B} d\lambda^{(1)} \frac{2n}{n^2+(\lambda^{(1)}-\lambda)^2} \, \epsilon_1^{(1)}(\lambda^{(1)}) \, ,$$

which is identical to what we found from zero temperature limit of the thermodynamics (see (3.17)), the expression for the excitation energy turns into

$$E_{ex} = \epsilon_n(\lambda_1^n) - \sum_{j=1}^{2N-2} \epsilon_1(\lambda_j^h) \, . \qquad (4.14)$$

By using the integral equations for $\epsilon_1$ and $\epsilon_1^{(1)}$ one can show that

$$\epsilon_n(\lambda) \equiv 0 \ .$$

Therefore the excitation energy is exactly the same as for an excitation with only $2N-2$ holes and no strings. The momentum for odd $N$ is given by

$$P = -\frac{2\pi}{L} \sum_{j=1}^{2(N-1)} I_j^h + \frac{2\pi I_\alpha^n}{L} \mod 2\pi \ ,$$

whereas for even $N$ we find

$$P = -\frac{2\pi}{L} \sum_{j=1}^{2(N-1)} I_j^h + \frac{2\pi I_\alpha^n}{L} + \frac{\pi}{L}(L \pm N_h)) \mod 2\pi \ .$$

As we are only considering excitations with a finite number of particles in the thermodynamic limit, the range of integers $I_\alpha^n$ is microscopic ($N - n \ll L$) and the contribution of $\frac{2\pi I_\alpha^n}{L}$ to the momentum goes to zero. Thus the momentum is identical to the momentum of an excitation with $2(N-1)$ holes only. The spin quantum numbers are found to be

$$S^z = N - n = S \ .$$

Excitations with an even number of holes and more than one string of length greater than 1 can be contructed in the same way as above. The result is that only the holes are dynamical, and the strings only affect the spin quantum number. This phenomenon is analogous to the situation for the Heisenberg magnet[14,35]. The interpretation of the spin-wave type excitations follows from the work of Faddeev and Takhtajan on the Heisenberg ferromagnet : in the spin sector of the $u(2|2)$ model there exists one two elementary excitations of spin $\pm\frac{1}{2}$ and charge zero (which are called spinons), but they are "confined" and only their scattering states (triplet, singlet, ...) are observable. As an example we consider the sectors of 2 and 4 spinons respectively. For 2 spinons there is the state with $N = 1$ (of the first type of excitations considered above), which has spin $S = 1 = S^z$ and is the lowest weight state of the triplet. The other state is the one with $N = 2 = n$ (of the second type of excitations considered above), which is the singlet state with $S = 0 = S^z$. Thus we obtain the correct irreducible representaions of the spin SU(2) for 2 particles.

In the sector with 4 spinons we have the state with $N = 2$, which is the lowest weight state of the symmetric $S = 2$ representation, three states with $N = 3$, $n = 2$ (as $|I_\alpha^3| \leq 1$), which lead to three $S = 1$ triplets, one singlet with $N = 3 = n$, and another singlet with $M_1 = N_h^{GS} + N_\downarrow^{GS} - 4$, $M_1^{(1)} = N_h^{GS}$, $M_2 = 2$ (which is a states with 4 holes and 2 strings of length 2). Altogether these are 16 states in the representation $\frac{1}{2} \otimes \frac{1}{2} \otimes \frac{1}{2} \otimes \frac{1}{2} = 2 \oplus 1 \oplus 1 \oplus 1 \oplus 0 \oplus 0$ as expected. This completes our analysis for the Bethe states in the spin-wave sector.

Each of the excitations constructed above is not only the lowest weight state of a spin SU(2) multiplet (as we took into account above), but of a $u(2|2)$ multiplet, the dimension of which is given by (1.7) or (1.8). Thus we can construct further excitations with different quantum numbers by applying the $u(2|2)$ raising operators. The energies of these excitations follow from the dispersion of the spin-waves and (1.9). We see that all excitations in the multiplet *except* the spin-SU(2) descendants acquire a gap with respect to the ground state for less than half filling. For example the state obtained by acting with $Q_{-1}$ on the spin singlet excitation constructed above, carries quantum numbers $S_z = \frac{1}{2}$ and charge $+e$ (the electron is taken to have charge $-e$) and has a gap of $\frac{U}{2} + \mu > 0$.

### Spectrum of Charge Excitations

Excitations connected with charge rearrangements that do not involve local electron pairs, are called holon-antiholon excitations and are completely analogous to the ones in the supersymmetric $t$-$J$ model. Therefore we will restrict ourselves to a brief discussion. A holon-antiholon excitation corresponds to rearrangements in the $\lambda^{(1)}$-distribution only. It is a particle-hole excitation, which is constructed by removing one of the integers $J_\gamma$ (corresponding to spectral parameter $\Lambda_h$) from the ground state distribution and placing it outside the "Fermi boundary" $\frac{\pm N_h^{GS}-1}{2}$ (which is the maximal integer taken in the ground state). We denote the spectral parameter corresponding to this newly "taken" integer by $\Lambda_p$. Taking the thermodynamic limit we obtain the following set of coupled integral equations for the densities

$$\rho_1(\lambda) = \frac{1}{2\pi}\frac{2}{1+\lambda^2} - \frac{1}{2\pi}\int_{-\infty}^{\infty} d\lambda' \frac{4}{4+(\lambda-\lambda')^2}\rho_1(\lambda')$$

$$+ \frac{1}{2\pi}\int_{-B}^{+B} d\lambda^{(1)} \frac{2}{1+(\lambda-\lambda^{(1)})^2}\rho_1^{(1)}(\lambda^{(1)}) + \frac{1}{2\pi L}\frac{2}{1+(\Lambda_p-\lambda)^2}$$

$$\rho_1^{(1)}(\lambda^{(1)}) = \frac{1}{2\pi}\int_{-\infty}^{\infty} d\lambda \frac{2}{1+(\lambda-\lambda^{(1)})^2}\rho_1(\lambda) - \frac{1}{L}\delta(\lambda^{(1)} - \Lambda_h) \ .$$

Defining $\psi_2 = -\delta(\Lambda - \Lambda_h)$ and $\psi_1 = \frac{1}{2\pi}\frac{2}{1+(\Lambda_p-\Lambda)^2}$, these equations can be written in matrix form (4.5), and following through the same steps as for the spin excitations above we find the excitation energy to be

$$E_{ex} = L \int_{-\infty}^{\infty} d\lambda\, \overset{(0)}{\varepsilon}_1(\lambda)\, \sigma_1(\lambda) + L \int_{-B}^{B} d\Lambda\, \overset{(0)}{\varepsilon}_2(\Lambda)\, \sigma_2(\Lambda) + \mu + \frac{U}{2}$$

$$= \sum_{\alpha=1}^{2} \int_{-a_\alpha}^{a_\alpha} d\mu_\alpha\, \varepsilon_\alpha(\mu_\alpha)\, \psi_\alpha + \mu + \frac{U}{2} = \epsilon_1^{(1)}(\Lambda_p) - \epsilon_1^{(1)}(\Lambda_h) \geq 0 \;. \quad (4.15)$$

Here $\sigma_i$ are the backflow distributions defined in (4.7). The contribution $\mu + \frac{U}{2}$ is due to the fact that $L \int_{-B}^{B} d\Lambda\, \rho_1^{(1)}(\Lambda) = N_h^{GS} - 1$, as the particle with $\Lambda_p$ is outside the Fermi boundary. The excitation energy clearly breaks up into a contribution from the hole $-\epsilon_1^{(1)}(\Lambda_h) \geq 0$ and from the particle $\epsilon_1^{(1)}(\Lambda_p) \geq 0$. The particle carries no spin and charge $+e$, whereas the hole carries no spin and charge $-e$. Adding the particle at the Fermi boundary $\pm B$ turns the two-parameter family into a one-parameter one which would describe only a hole. At half filling $B = 0$ the complete $\lambda^{(1)}$ sea is empty and one can make neither particle-hole excitations nor an elementary particle excitations[26] with no spin (one gets scattering states of holons and spinons). Once again one can construct the complete $u(2|2)$ multiplet from this excitation by acting with the raising operators. As the charge excitations all are spin singlets, the corresponding multiplet is of dimension $16(N_h + 1)$, where $N_h + 1$ is the dimension of the $\eta$-SU(2). However all these excitations have a gap for $\mu < 0$. For $\mu = 0$ the localonic excitations become gapless (see below).

<div align="center"><em>Localonic Spectrum</em></div>

Let us now turn to the study of excitations involving local electron pairs. Such excitations are not present in the supersymmetric $t$-$J$ model. We first turn to excitations involving $\lambda^{(2)}$-strings. We consider the following situation

$$M_1 = N_h^{GS} + N_\downarrow^{GS},\; M_n = 0\; \forall n \geq 2 \;,$$
$$M_1^{(1)} = N_h^{GS},\; M_p^{(2)} = 1\;,\; M_2^{(1)} = 0 = M_q^{(2)}\; \forall q \neq p \;.$$

This excitation corresponds to replacing $p$ holes (empty sites) in the ground state with a $\lambda^{(2)}$ string of length $p$. The excitation is spinless $S = 0$ and carries charge $-2pe$. It describes an excitation of $p$ bound local electron pairs. The integers $I_\alpha^1$ are distributed exactly like in the ground state, whereas the $J_\gamma$ jump from half-odd integers to integers. The allowed range of integers for the $\lambda^{(2)}$ string follows from (2.12) and is found to be $|N_1^p| \leq \frac{1}{2}(N_h^{GS} - 2p)$. The spectral parameter describing the center of the $p$-string is determined by the integer $N_1^p$ via the equation

$$2\pi N_1^p = \sum_{\gamma=1}^{M_1^{(1)}} \theta\left(\frac{\kappa - \lambda_\gamma^{(1)}}{p}\right) \;. \quad (4.16)$$

The integral equations for the $\lambda$ and $\lambda^{(1)}$ distribution functions are found to be

$$\rho_1(\lambda) = \frac{1}{2\pi}\frac{2}{1+\lambda^2} - \frac{1}{2\pi}\int_{-\infty}^{\infty} d\lambda' \frac{4}{4+(\lambda-\lambda')^2}\,\rho_1(\lambda')$$
$$+ \frac{1}{2\pi}\int_{-B}^{+B} d\lambda^{(1)} \frac{2}{1+(\lambda-\lambda^{(1)})^2}\,\rho_1^{(1)}(\lambda^{(1)}) \qquad (4.17)$$
$$\rho_1^{(1)}(\lambda^{(1)}) = \frac{1}{2\pi}\int_{-\infty}^{\infty} d\lambda \frac{2}{1+(\lambda-\lambda^{(1)})^2}\,\rho_1(\lambda) - \frac{1}{2\pi L}\left(\frac{2p}{p^2+(\lambda^{(1)}-\kappa)^2}\right).$$

If we define

$$\psi_1 = 0\,,\quad \psi_2 = -\frac{1}{2\pi}\frac{2p}{p^2+(\lambda^{(1)}-\kappa)^2}$$

the backflow-distribution once again obey a matrix integral equation of type (4.8). The excitation energy follows from (2.13) to be

$$E(\kappa) = E(B) - E_{GS}(B_0) = -2p\mu + \sum_{\alpha=1}^{2}\int_{-a_\alpha}^{a_\alpha} d\mu_\alpha\,\varepsilon_\alpha(\mu_\alpha)\,\psi_\alpha$$
$$= -2p\mu - \frac{1}{2\pi}\int_{-B}^{B} d\Lambda\,\frac{2p}{p^2+(\Lambda-\kappa)^2}\,\epsilon_1^{(1)}(\Lambda)\ .$$

Using the expression for the dressed energy of a $\lambda^{(2)}$-string of length $p$ given by (3.17) this can be written in the form

$$E(\kappa) = \epsilon_p^{(2)}(\kappa)\ \geq -2\mu p > 0.$$

We see that this kind of excitation has a gap for $\mu < 0$, which is the case we are currently investigating. This gap goes to zero as $\mu$ approaches zero.

The dressed momentum of $\lambda^{(2)}$-string excitations can be evaluated using (2.14) and (4.16)

$$P(\kappa) = \frac{2\pi}{L}\left(\pm\frac{M_1^{(1)}}{2} - N_1^p\right) = \pm\pi\frac{N_h^{GS}}{L} - \int_{-B}^{B} d\Lambda\ \theta(\tfrac{\kappa-\Lambda}{p})\,\rho_1^{(1)}(\Lambda)\ .$$

Note that the $\pm$ sign is due to the fact that there exist two possibilities of distributing the integers $J_\gamma = J_\gamma^{GS} \pm \frac{1}{2}$. Let us concentrate on the case with the plus sign (the minus case can be treated in the same way). In the limit $\kappa \to \infty$, $\kappa \gg B$, the momentum goes to zero as

$$P(\kappa) \xrightarrow[\kappa\to\infty]{} \int_{-B}^{B} d\Lambda\ \frac{2p}{\kappa-\Lambda}\,\rho_1^{(1)}(\Lambda) \sim \frac{2p}{\kappa}D_h^{GS}\ ,$$

where $D_h^{GS} = \frac{N_h^{GS}}{L}$ is the density of holes in the ground state. We observe that zero momentum corresponds to infinite $\kappa$. The excitation energy to leading order in $\frac{1}{\kappa}$ is

$$E(\kappa) \xrightarrow[\kappa \to \infty]{} -2p\mu - \frac{2p}{\kappa^2} \frac{1}{2\pi} \int_{-B}^{B} d\Lambda \; \epsilon_1^{(1)}(\Lambda).$$

Combining the asymtotic expressions for energy and momentum, we find a quadratic dispersion relation with a gap proportional to the magnitude of the chemical potential for small momenta

$$E(P) = -2p\mu + \frac{P^2}{2\mathcal{M}} + \mathcal{O}(P^3),$$

where $\mathcal{M}$ plays the role of a mass and in our approximation is given by

$$\mathcal{M} = \frac{2p\pi (D_h^{GS})^2}{-\int_{-B}^{B} d\Lambda \; \epsilon_1^{(1)}(\Lambda)} > 0 \; .$$

Here $\int_{-B}^{B} d\Lambda \; \epsilon_1^{(1)}(\Lambda)$ is the energy of the particles in the $\lambda^{(1)}$-ground state distribution.

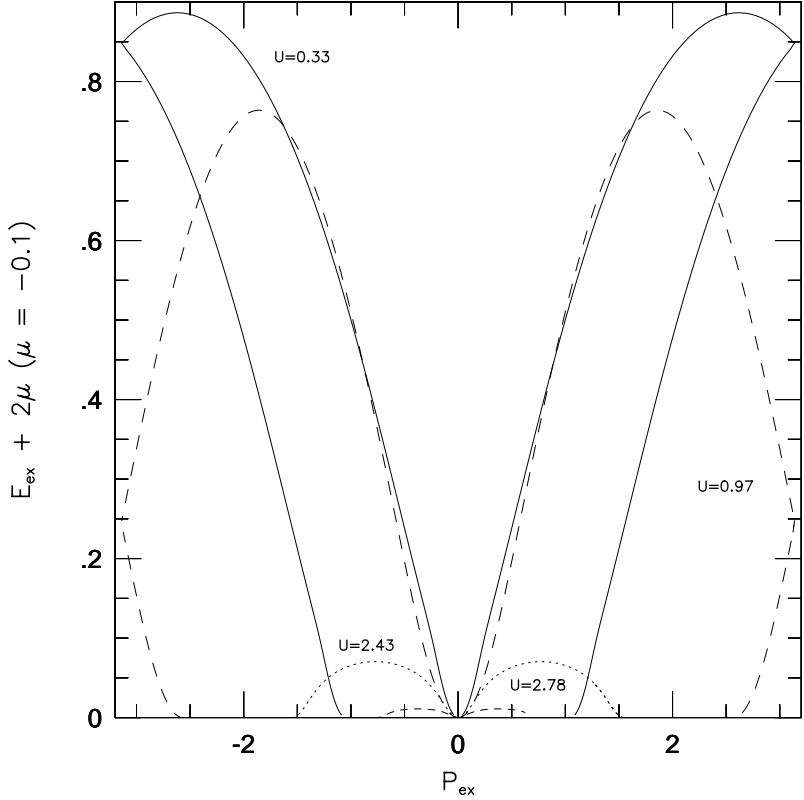

Figure 3: Dispersion relations for a $\lambda^{(2)}$-string of length 1 for $\mu = -0.1$ and various values of $U$.

Figure 3 displays the dispersion relations for a localonic excitation with $p = 1$, i.e. for a $\lambda^{(2)}$-string of length 1, for various values of $U$ and fixed chemical potential $\mu = -0.1$. The gap for this particular excitation is $-2\mu$, and we have displayed the whole range of momentum by taking in account both signs for the shift of the integers $J_\gamma$ (the $+$ ($-$) sign leads to positive (negative) momentum, of the same type of excitation). From Figure 3 we see that the allowed range of momentum tends to zero as $\mu + \frac{U}{2}$ approaches $2\ln(2)$, which corresponds to the ground state filled with single electrons (half-filled $t$-$J$ ground state). On the other hand, as $U$ goes to zero we observe an "extended Brillouin zone" - the single particle excitations appear multivalued in the interval $[-\pi, \pi]$, and only appear as single valued in an extended Brillouin zone scheme in a larger interval. This is a well know phenomenon, see e.g. [36]. Excitations of $\lambda^{(2)}$-strings of length 1 with zero momentum correspond to the action of the symmetry generator $\eta^\dagger$. As zero momentum corresponds to infinite spectral parameter $\lambda^{(2)}$ there is no Pauli-principle for local electron pairs with zero momentum (which is in perfect agreement with the $u(2|2)$ symmetry of the model).

It is rather instructive to investigate the zero-density limit in some detail. This limit corresponds to taking $\mu + \frac{U}{2}$ to zero, which is equivalent to $B \to \infty$. In this limit it is possible to solve the coupled integral equations (3.18) analytically with the result $\epsilon_1^{(1)}(\Lambda) = -\frac{8}{4+\Lambda^2}$. Thus the excitation energy for a single localon ($\lambda^{(2)}$-string of length 1) follows to be $E(\lambda^{(2)}) = \epsilon_1^{(2)}(\lambda^{(2)}) = -2\mu + \frac{12}{9+(\lambda^{(2)})^2}$. The state is described by the quantum numbers $N_\downarrow = 1 = N_\uparrow$, $N_l = 1$[9]. Naively one might have expected to arrive at the state with one local pair over the empty vacuum. Although this is indeed an eigenstate of the hamiltonian, it is *not* given by the Bethe Ansatz in the BBFF representation, as was shown in [10]. Only lowest weight states of the $u(2|2)$ algebra are Bethe Ansatz states, and the one-localon state is a $u(2|2)$ descendant of the state with one spin down (in the BBFF representation).

### *Electronic Spectrum*

The other type of excitation involving local electron pairs is the $\lambda^{(1)} - \lambda^{(2)}$-string excitation. It will turn out however, that is does not describe a localonic excitation but rather carries the quantum numbers of an electron. It is constructed by choosing

$$M_1 = N_h^{GS} + N_\downarrow^{GS}, \; M_n = 0 \;\; \forall n \geq 2 \;,$$
$$M_1^{(1)} = N_h^{GS} - 1, \; M_2^{(1)} = 1 \;, \; M_q^{(2)} = 0 \;\; \forall q \;.$$

Like for the ground state all vacancies for the integers $I_\alpha^1$ are filled symmetrically around zero. The $J_\gamma$'s change from half-odd integer to integer. However, the number $M_1^{(1)}$ of integers $J_\gamma$ also decreases by one as compared to the ground state, so that they are again distributed symmetrically around zero. The allowed range of integers for the $\lambda^{(1)} - \lambda^{(2)}$-string follows from (2.12) to be equal to $|K_\delta| \leq \frac{1}{2}(N_\downarrow^{GS} - 1)$. The rapidity $\kappa$ describing the center of the $\lambda^{(1)} - \lambda^{(2)}$ string is determined by its integer via

$$2\pi K_\delta = \sum_{\alpha=1}^{M_1} \theta\left(\frac{\kappa - \lambda_\alpha}{2}\right) - \sum_{\gamma=1}^{M_1^{(1)}} \theta\left(\kappa - \lambda_\gamma^{(1)}\right) . \tag{4.18}$$

---

[9]In the FFBB representation (for which the structure of the string hypothesis is the same as for the BBFF representation, see below) it is given by the state with one $\lambda$ string of length 3, and one $\lambda^{(1)}$-$\lambda^{(2)}$ string.

Once again the backflow distributions obey a matrix integral equation of the form (4.8) where now the functions $\psi$ are given by

$$\psi_1 = \frac{1}{2\pi} \frac{4}{4 + (\Lambda - \kappa)^2} , \quad \psi_2 = -\frac{1}{2\pi} \frac{2}{1 + (\Lambda - \kappa)^2} .$$

In terms of dressed energies the excitation energy is given by

$$E(\kappa) = U + \frac{1}{2\pi} \int_{-\infty}^{\infty} d\lambda \, \epsilon_1(\lambda) \frac{4}{4 + (\lambda - \kappa)^2} - \frac{1}{2\pi} \int_{-B}^{B} d\Lambda \, \epsilon_1^{(1)}(\Lambda) \frac{2}{1 + (\Lambda - \kappa)^2} .$$

Comparing this expression with the zero temperature limit of the TBA (3.17) we see that this is precisely equal to the dressed energy of a $\lambda^{(1)}$-$\lambda^{(2)}$ string, i.e.

$$E(\kappa) \equiv \epsilon_2^{(1)}(\kappa) .$$

The dressed momentum can be determined using (2.14) and (4.18)

$$P(\kappa) = \int_{-\infty}^{\infty} d\lambda \, \rho_1(\lambda) \, \theta\left(\frac{\kappa - \lambda}{2}\right) - \int_{-B}^{B} d\Lambda \, \rho_1^{(1)}(\Lambda) \, \theta(\kappa - \Lambda) .$$

The quantum numbers carried by this type of excitation are $S = \frac{1}{2} = S^z$ and charge $-e$, so that it may be interpreted as an *electron*. The dispersion relation is plotted below in Figure 4 for a fixed chemical potential of $\mu = -0.1$ and various values of $U$. Clearly there is always a gap, which tends to zero for $\mu + \frac{U}{2} \to 0$. In this limit however the allowed range of momenta also goes to zero, so that the state cannot be excited. On the other hand, when $\mu + \frac{U}{2} \to 2\ln(2)$ (when the ground state is equal to the half filled ground state of the supersymmetric $t$-$J$ model) the allowed range of momentum is $[-\pi, \pi]$.

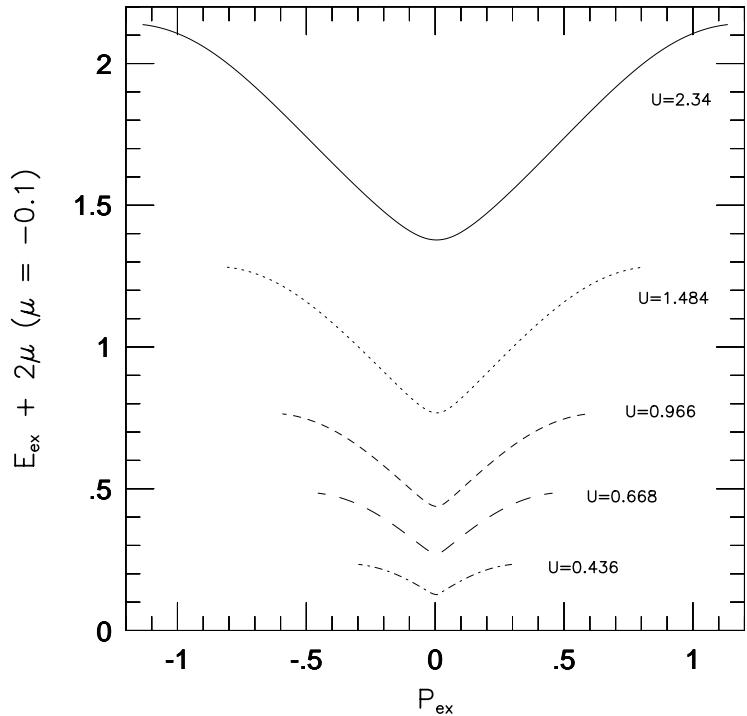

Figure 4: Dispersion relation for a $\lambda^{(1)}$-$\lambda^{(2)}$ string.

For small momenta the dispersion relation is quadratic and has a gap

$$E(P) = U - \epsilon_1(0) - 4 + \frac{P^2}{2(\frac{1-\pi\ \rho_1(0)}{4+\frac{\pi^2}{8}})} + \mathcal{O}(P^3)\ .$$

Note that is the region under consideration $U > 0$ ($\epsilon_1(0) < 0$ always), and the gap is positive. For momenta close to the Fermi momentum $k_F$ the dispersion is of the form

$$E(P) \sim U - (P - k_F)^2 + \mathcal{O}((P - k_F)^3)\ .$$

This shows that the gap approaches $U$ at the Fermi boundary, which is in perfect agreement with Figure 4 (note that in Figure 4 we have subtracted $-2\mu$ from the energy).

This concludes our analysis in region III. In summary we find the *gapless* excitation spectrum to be the one of the supersymmetric *t-J* model. However there exists a large number of new excitations with a gap.

## 4.2. Excitations in Region I

We have shown in [10] that the ground state in region I, that is for zero chemical potential and attractive on-site interaction $U < 0$, is of the form $(\eta^\dagger)^N|0\rangle$, where $2N$ is the total number of electrons (note that we consider the ground state only for even total numbers of electrons, because it is otherwise trivially degenerate). The analysis of the excitation spectrum in this sector is most easily performed in the FFBB representation (see [10]). The reference state in this representation is the empty vacuum $|0\rangle$, which is much "closer" to the ground state than the reference state of the BBFF representation used above. The ground state is of a similar nature as the purely ferromagnetic ground state in the spin-$\frac{1}{2}$ Heisenberg XXX model, as it is also an $SU(2)$ descendant of the bare (empty) vaccum $|0\rangle$. Due to this nature of the ground state one would not expect the excitations to get a nontrivial dressing from the ground state electrons as is the case for antiferromagnetic systems. This is indeed the case as we will now proceed to show[10]. In [10] we have constructed the Bethe Ansatz eigenstates of the hamiltonian, which are of the form

$$\prod_{i=1}^{m} C_{a_i}(\lambda_i)|0\rangle F^{a_m\ldots a_1} . \tag{4.19}$$

Here the $C_{a_i}(\lambda_i)$ are creation operators with respect to the vacuum $|0\rangle$. A complete set of eigenstates can be obtained by acting with the $u(2|2)$ raising operators on these states. We furthermore evaluated all commutators of the generators of the $u(2|2)$ algebra with the creation operators $C$. As a special case we find

$$\left[\eta^\dagger, C_a(\lambda)\right] = 0 .$$

This implies that

$$(\eta^\dagger)^n \prod_{i=1}^{m} C_{a_i}(\lambda_i)|0\rangle F^{a_m\ldots a_1} = \prod_{i=1}^{m} C_{a_i}(\lambda_i)(\eta^\dagger)^n|0\rangle F^{a_m\ldots a_1} . \tag{4.20}$$

Recalling that a complete set of states is obtained by acting with the $u(2|2)$ raising operators of the states (4.19) we arrive at the following expression for a *complete set of excitations over the ground state* $(\eta^\dagger)^N|0\rangle$

$$u(2|2) \times \prod_{i=1}^{m} C_{a_i}(\lambda_i)(\eta^\dagger)^N|0\rangle F^{a_m\ldots a_1} ,$$

---

[10] Note that this would not be obvious if we operated in the BBFF representation.

where $u(2|2) \times$ denotes action by all raising operators of $u(2|2)$ and $\eta$. Now we see that due to (4.20) the only nontrivial part of the energies comes from the Bethe states, as action with the $u(2|2)$ generators just shifts the energies according to (1.9) (and the hermitean conjugate equations). Let us now look at some excitations in more detail. It was shown in [10] that the Bethe equations for the BBFF and the FFBB representations are identical up to the substitutions $N_h \leftrightarrow N_\uparrow$ and $N_l \leftrightarrow N_\downarrow$. Thus the logarithmic form of the Bethe equations and the boundaries for the integers derived in section 2 for the BBFF representation also hold in the FFBB representation up to these substitutions. Energy and momentum in the FFBB representation are given by

$$E(\{\lambda_\alpha^n\}) = \sum_{n=1}^{\infty} \sum_{\alpha=1}^{M_n} \left( \frac{4n}{(\lambda_\alpha^n)^2 + n^2} - 2\mu n \right) - L(1 - \frac{U}{4})$$

$$+ M_1^{(1)}(\mu - \frac{U}{2} + h) + M_2^{(1)}(2\mu - U) - \sum_{p=1}^{\infty} \sum_{\nu=1}^{M_p^{(2)}} 2hp \quad , \qquad (4.21)$$

$$P(\{\lambda_\alpha^n\}) = \sum_{n=1}^{\infty} \sum_{\alpha=1}^{M_n} \pi - 2 \arctan\left(\frac{\lambda_\alpha^n}{n}\right)) \quad ,$$

where now

$$M := \sum_n n M_n = N_\downarrow + N_\uparrow + N_l \quad ,$$

$$M^{(1)} := M_1^{(1)} + 2 M_2^{(1)} = N_\downarrow + N_\uparrow \quad ,$$

$$M^{(2)} := M_2^{(1)} + \sum_p p M_p^{(2)} = N_\downarrow \quad .$$

An excitation of a bound state of $n$ local electron pairs over the *bare* vacuum $|0\rangle$ (and by the above also over the true ground state $(\eta^\dagger)^N |0\rangle$) is easy to construct as it only involves one spectral parameter of the zeroth level Bethe equations. Picking $M_n = 1 = M$, $M_1^{(1)} = 0 = M_2^{(1)}$, and $M_p^{(2)} = 0$, we find the following dispersion relation (for $h = 0$ and $\mu = 0$)

$$E(P) = \frac{4}{n}\sin^2\left(\frac{P}{2}\right) \quad ,$$

where $P$ is the momentum of the excitation with respect to the ground state. The quantum numbers are $S = 0$ and charge $-2n$. By means of the supersymmetry we can construct electronic excitations as follows : consider an excitation with a single local pair over the ground state as constructed above ($n = 1$). Denote the wave function of this state by $|loc\rangle$. The state $\tilde{Q}_\sigma |loc\rangle$ has a dispersion relation $E(P) = 4\sin^2\left(\frac{P}{2}\right) - \frac{U}{2}$, spin $-\sigma$ and charge $-e$, i.e. describes an electronic excitation over the ground state. Acting with $Q_\sigma^\dagger$ on $|loc\rangle$ leads to a "three particle excitation" (one single electron and one local pair) with the same dispersion as $\tilde{Q}_\sigma|loc\rangle$. Of course there is an infinite number of other types of excitations, e.g. $\lambda^{(2)}$-string excitations of any length. Our discussion above shows how all of these can be constructed in principle from the Bethe Ansatz and the $u(2|2)$ supersymmetry.

### 4.3. Excitations in Region II

At $\mu = 0$ the compressibility becomes infinite, and the ground state density is no longer determined by the value of the chemical potential[7]. It is therfore necessary to work in the canonical ensemble and fix the density of particles $D = \frac{N_e + 2N_l}{L}$. The ground state in region II (that is for $\mu = 0$, $0 \leq U < 4\ln(2)$) is of the form $(\eta^\dagger)^n |stJ\rangle$, where $|stJ\rangle$ is the ground state of the supersymmetric $t$-$J$ model at chemical potential $\mu_{tJ} = \frac{U}{2}$, and $n$ is determined by the density imposed. This region is of particular interest, as the ground state in this region exhibits off-diagonal long-range order[7] and is thus superconducting. The ground state can also be determined from the zero temperature limit of the thermodynamic equations: for $\mu = 0$ the dressed energies of localonic excitations become gapless, and due to the absence of a Pauli principle for local pairs at zero momentum there is an infinite number of states with the same energy as the ground state obtained by taking the limit $\mu \to 0$ from the left. Thus there is no unique ground state in the grand canonical ensemble, but by imposing a fixed density $D$ we can single out a unique ground state, which is of the above form. The excitation spectrum now can be easily constructed by combining the results of sections 4.1. and 4.2.. By taking the limit $\mu \to 0$ of the spectrum determined in 4.1. we obtain the excitations over the state $|stJ\rangle$, which by (4.20) readily translates into the excitation spectrum over the ground state $(\eta^\dagger)^n|stJ\rangle$. The result for $0 \leq U < 4\ln(2)$ is as follows: the spin-wave and charge excitations are like for $\mu < 0$ (gapless, linear dispersion around their Fermi surfaces). The localonic excitations are now *gapless*, but still have a quadratic dispersion. The electronic excitation still has a gap proportional to $U$. The quadratic behaviour of the gapless dispersion of the localons clearly indicates non-conformal properties of the model. The methods of conformal field theory[37-42] can therefore not be applied directly to evaluate the critical exponents. However, by using the algebra of the $u(2|2)$ generators we have succeeded to express correlators in the $u(2|2)$ model is terms of correlators in the supersymmetric $t$-$J$ model. This enables us to evaluate the critical exponents using the conformal field theory results for the $t$-$J$ model[43,44].

Like for $\mu < 0$ every Bethe Ansatz excitation provides the lowest weight state of a whole $u(2|2)$ multiplet, the structure of which is given by (1.7) and (1.9).
For $U \geq 4\ln(2)$ $|stJ\rangle$ is the half-filled ground state of the supersymmetric $t$-$J$ model, for which the excitation spectrum is rather different as compared to less than half-filling[26]. We leave the discussion of the region for a future publication.

### 4.3. Excitations in Region V

Excitations in region V can be constructed the same way as in region I. The main difference is that the chemical potential in region V is not zero, and that only the action of the $u(2|2)$ *raising* operators has to be taken into account.

ACKNOWLEDGEMENTS:

It is a pleasure to thank Kareljan Schoutens for many useful discussions and his help in deriving the string hypothesis for this model.